\documentclass{article}
\usepackage{iclr2027_conference,times}
\usepackage{latexsym}
\usepackage[T1]{fontenc}
\usepackage[utf8]{inputenc}
\usepackage{microtype}
\usepackage{graphicx}
\usepackage{booktabs}
\usepackage{amsmath}
\usepackage{amssymb}
\usepackage[table]{xcolor}   %

\definecolor{dvneg2}{HTML}{E8A0A0}
\definecolor{dvneg1}{HTML}{F6CFCF}
\definecolor{dvzero}{HTML}{F0EFEC}
\definecolor{dvpos1}{HTML}{CDE2FB}
\definecolor{dvpos2}{HTML}{9EC5F4}
\definecolor{dvpos3}{HTML}{5598E7}
\usepackage{float}
\usepackage{tikz}
\usetikzlibrary{positioning,arrows.meta,decorations.pathreplacing}
\usepackage[colorlinks=true,linkcolor=blue!50!black,citecolor=blue!50!black,urlcolor=blue!50!black]{hyperref}
\usepackage{url}

\iclrfinalcopy

\newcommand{\consistpos}{\textbf{stable} $+$}
\newcommand{\consistneg}{\textbf{stable} $-$}
\newcommand{\flips}{\textcolor{red}{sign flips}}

\title{What Transfers from a VLM Teacher?\\Comparing Supervision Signals\\for Visual Document
Retrieval}

\author{
Saba Sturua \hspace{2em} Han Xiao \\[2pt]
\normalfont Jina AI by Elastic \\
\normalfont 33 New Montgomery Street, San Francisco, CA 94105, USA \\
\normalfont \texttt{research@jina.ai}
}

\begin{document}
\maketitle
\lhead{Preprint.}

\begin{abstract}
Visual document retrievers are trained contrastively: each query is matched to one page labelled
relevant --- the positive --- and pushed away from negatives, pages presumed irrelevant. Recent
methods distil a vision--language model (VLM) teacher into the retriever by enriching that
positive, transferring the teacher's attention over it or a description of it. We ask whether the
teacher is better spent on the other side, judging the candidates the retriever mines as
negatives, which the label says nothing about. With student, data, optimizer and evaluation fixed,
teacher-judged hard negatives and score distillation raise ViDoRe~v2 nDCG@5 from $55.2$ to $62.6$
and $63.0$; description alignment, as adapted here, gains $2.6$ points and attention grounding nothing measurable.
Against teacher-free rules that select four candidates from the same mined pool at identical
training compute, the best of which is
the positive-aware threshold current systems use, the teacher's judgement adds $4.1$ points on v2
and $1.7$ on v3. This is consistent with how incomplete the labels are. Annotators judge about two
of a query's four top-ranked mined candidates relevant, none of them labelled, so training pushes
the retriever away from relevant pages treated as negatives. What reaches the student is coarse: under a greedily decoded $0$--$100$ rating prompt, $82\%$ of the
teacher's ratings come back at one end of the scale or the other, and a relevant/irrelevant
partition keeps most of the
distillation gain. A ten-annotator audit places the teacher within the range of variation among
human annotators, and finds it reliable where a query has a single determinate
answer. We release the code, the teacher's $3.3$M judgements and page descriptions, the mined pools, the human audit and the trained adapters at \url{https://github.com/elastic/vdr-teacher-signals}.

\end{abstract}

\section{Introduction}
\label{sec:intro}
Visual document retrieval encodes each page as an image rather than extracting the text in it,
which preserves what layout carries: the structure of a table, or a
label inside a figure. DSE \citep{dse} established the approach, and a dominant family builds on ColPali and its descendants \citep{colpali}, which keep per-patch embeddings and score relevance
by ColBERT-style late interaction \citep{colbert}. Training is contrastive: each query is compared
against one page labelled relevant --- the positive --- and against unlabelled others treated as negatives.

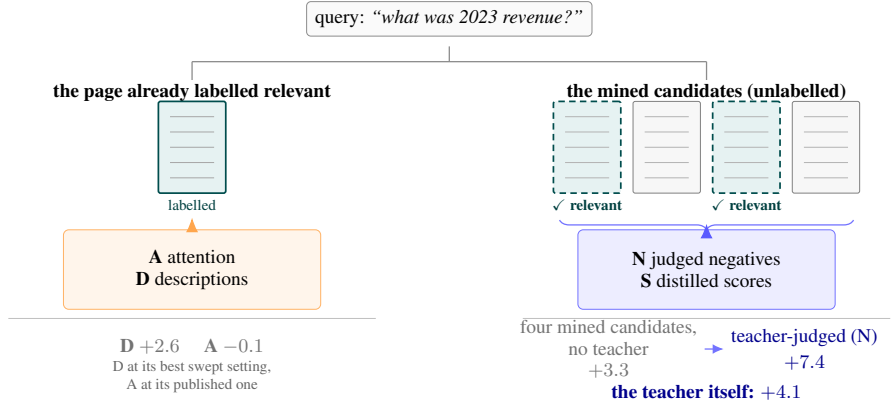
\begin{figure*}[t]
\centering \resizebox{0.84\textwidth}{!}{\begin{tikzpicture}[
  font=\small,
  x=1mm, y=1mm,
  page/.style   = {draw=black!50, fill=black!3, rounded corners=1pt,
                   minimum width=11mm, minimum height=15mm, inner sep=0pt},
  gold/.style   = {page, fill=teal!10, draw=teal!60!black, thick},
  relv/.style   = {page, fill=teal!10, draw=teal!60!black, thick, densely dashed},
  hdr/.style    = {font=\small\bfseries, align=center},
  sig/.style    = {draw, rounded corners=2.5pt, align=center, inner sep=4pt,
                   minimum width=42mm, minimum height=13mm, font=\small},
  describe/.style = {sig, fill=orange!8,  draw=orange!60},
  judge/.style    = {sig, fill=blue!7,    draw=blue!55},
  note/.style    = {font=\scriptsize, text=black!60, align=center},
  verdict/.style  = {font=\small\bfseries, align=center},
  up/.style     = {-{Latex[length=2.4mm,width=2mm]}, line width=0.6pt},
  split/.style  = {draw=black!45, semithick},
]

\def\colL{-42}   %
\def\colR{42}    %

\node[draw=black!45, rounded corners=2pt, fill=black!2, inner sep=4pt, font=\small]
  (q) at (0,34) {query: \emph{``what was 2023 revenue?''}};

\draw[split] (0,29.5) -- (0,27);
\draw[split] (\colL,27) -- (\colR,27);
\draw[split] (\colL,27) -- (\colL,24);
\draw[split] (\colR,27) -- (\colR,24);

\node[hdr] at (\colL,22) {the page already labelled relevant};
\node[hdr] at (\colR,22) {the mined candidates (unlabelled)};

\node[gold] (g) at (\colL,13) {};
\foreach \yy in {18,15.5,13,10.5,8}{\draw[black!35] (\colL-3.6,\yy) -- (\colL+3.6,\yy);}
\node[font=\scriptsize, text=teal!50!black] at (\colL,3.5) {labelled};

\foreach \i/\dx/\st in {1/-19.5/relv, 2/-6.5/page, 3/6.5/relv, 4/19.5/page}{
  \node[\st] (c\i) at (\colR+\dx,13) {};
  \foreach \yy in {18,15.5,13,10.5,8}{\draw[black!30] (\colR+\dx-3.6,\yy) -- (\colR+\dx+3.6,\yy);}
}
\foreach \dx in {-19.5,6.5}{
  \node[font=\scriptsize\bfseries, text=teal!50!black] at (\colR+\dx,3.5) {$\checkmark$ relevant};
}

\node[describe] (ad) at (\colL,-7) {\textbf{A} attention\\\textbf{D} descriptions};
\node[judge]    (ns) at (\colR,-7) {\textbf{N} judged negatives\\\textbf{S} distilled scores};

\draw[up, orange!70] (\colL,-0.5) -- (\colL,1.5);
\draw[decorate, decoration={brace, amplitude=1.6mm, mirror, raise=0.4mm},
      draw=blue!55, semithick] (\colR-24,1.5) -- (\colR+24,1.5);
\draw[up, blue!65]   (\colR,-0.5) -- (\colR,0.5);

\draw[black!25] (\colL-30,-15) -- (\colL+30,-15);
\draw[black!25] (\colR-30,-15) -- (\colR+30,-15);

\node[verdict, text=black!55] at (\colL,-19.5) {D $+2.6$ \quad A $-0.1$};
\node[note] at (\colL,-24.5) {D at its best swept setting,\\A at its published one};

\node[font=\small, text=black!55, align=center] (nt) at (\colR-16,-20)
  {four mined candidates,\\no teacher\\\textbf{$+3.3$}};
\node[font=\small, text=blue!55!black, align=center] (wt) at (\colR+16,-20)
  {teacher-judged (N)\\[2pt]\textbf{$+7.4$}};
\draw[-{Latex[length=1.8mm,width=1.6mm]}, blue!55, semithick] (nt.east) -- (wt.west);
\node[font=\small\bfseries, text=blue!55!black, align=center] at (\colR,-27)
  {the teacher itself: $+4.1$};

\end{tikzpicture}}
\caption{Where a teacher can be spent: (\textit{left}) on the page already labelled relevant
(A, D), or (\textit{right}) on the retriever's unlabelled mined candidates (N, S;
\S\ref{sec:labels}). Numbers are ViDoRe~v2 nDCG@5 changes against the same model trained without a teacher (Table~\ref{tab:main}). On the candidate side, the teacher's own contribution is the gap to a teacher-free rule picking four candidates from the same pool
(Table~\ref{tab:contributes}).}
\label{fig:schematic}
\end{figure*}

\paragraph{The labels are incomplete.} Training treats that one label as if every other page were irrelevant, and in the standard ColPali training corpus this is often wrong. Its queries come from
document-VQA datasets and synthetic collections, each written or generated against a single page,
so ``What is the date?'' can be
answered by any page bearing a date while only one of them is marked correct. Hard-negative mining
surfaces exactly those pages, so training pushes the retriever away from pages that answer the query. In a ten-annotator audit, about
two of a query's four top-ranked mined candidates are judged relevant, none of them labelled (\S\ref{sec:labels}).

\paragraph{Where teachers are spent today.}
Two recent methods distil a larger vision--language model (VLM) into a visual document retriever, and
both enrich the labelled page: AGREE \citep{agree} transfers the teacher's \emph{attention} over
it, ReAlign \citep{realign} a generated \emph{description} of its contents. Text retrieval and
recent multimodal embedding work instead spend a judge on the mined candidates, and leading visual
document retrievers filter those candidates with a teacher-free similarity threshold, some also distilling a reranker's scores over them (\S\ref{sec:related}). These two uses of a teacher have not been compared under controlled
conditions, and how much of a judge's output survives as a training target has received little study. We do both in one tightly controlled setting --- one student, one corpus, and one teacher
for every signal but attention, which uses the smaller teacher its authors recommend
(\S\ref{sec:setup}) --- trading breadth for control.

\paragraph{Contributions.}
\begin{itemize}\itemsep0pt \parskip0pt \topsep2pt
\item \textbf{A controlled comparison} of four ways to spend a VLM teacher on a visual document
retriever --- teacher-judged hard negatives (N) and score distillation (S), which judge mined
candidates, and attention grounding (A) and description alignment (D), which enrich the labelled
page --- with student, data, order, optimizer and harness fixed, seed-matched baselines, and
teacher-free controls at identical training compute (Figure~\ref{fig:schematic}). Judging candidates
improves ViDoRe~v2 by $7.4$--$7.8$ nDCG@5 against $2.6$ and ${\approx}0$, and the teacher's own
judgement adds $+4.1$ on v2 and $+1.7$ on v3 (\S\ref{sec:results}--\ref{sec:contributes}).
\item \textbf{Evidence that the training set's single-positive labels are substantially
incomplete}, from three directions that agree: how its queries were written, what the teacher
rates relevant, and what annotators judge relevant (\S\ref{sec:labels}).
\item \textbf{An analysis of what the teacher transmits}: under a greedily decoded $0$--$100$ rating the signal is near-binary, a relevant/irrelevant partition keeps most of the distillation gain, and, in a
post-hoc analysis, constant-target rows act as a uniformity penalty (\S\ref{sec:anatomy}).
\item \textbf{A ten-annotator audit} of $400$ judgements that places the teacher within the range
of human agreement and locates where it is least reliable: queries with no single determinate answer
(\S\ref{sec:validation}).
\item \textbf{Resources we release}: the code, the teacher's $3.3$M relevance judgements and page
descriptions, the frozen mined pools, the $400$ human judgements, and trained adapters for five arms (the baseline, N, S, the strongest teacher-free control and the best-scoring variant of S), so the judged negatives and score targets can be reused without a 72B teacher.
\end{itemize}

\section{Related work}
\label{sec:related}

\paragraph{Visual document retrieval.} Since DSE \citep{dse} and ColPali \citep{colpali}, work has largely varied the backbone or the deployment constraints \citep{modernvbert,gme,jinav4,visrag,vdocrag}, while the supervision has stayed one query, one labelled page, and negatives drawn from the batch or the retriever's own mining. Where current visual document retrievers filter mined negatives, they do so without a teacher, with a positive-aware similarity threshold \citep{colembedv2,qwen3vlemb,kovre} of the kind introduced by \citet{nvretriever}, the control \S\ref{sec:contributes} measures the teacher against; two of them also distil a reranker's scores over mined candidates, a form of S \citep{qwen3vlemb,kovre}. From the evaluation side, ViDoRe~v2's authors criticise extractive, single-page queries \citep{vidorev2}, the benchmark-side view of the incompleteness we measure in \S\ref{sec:labels}.

\paragraph{Distillation into retrievers.} In text retrieval, a cross-encoder teacher's scores \citep{kd,monot5} supervise a cheaper bi-encoder through a margin objective \citep{marginmse} or soft scores over a candidate list \citep{colbertv2}. Two recent methods bring it to visual document retrieval, both with a positive-level target: AGREE
\citep{agree} transfers the teacher's attention over the page and ReAlign \citep{realign} a
generated description of it.

\paragraph{Structural problems with contrastive training data.} Mined hard negatives are routine in text retrieval \citep{dpr,ance}, but mining surfaces unlabelled positives that are then trained against. RocketQA filters them with a stronger model \citep{rocketqa}, sampling slightly easier candidates helps \citep{simans}, and contrastive learning debiases the objective \citep{chuang2020,robinson2021}, as neural ranking models pooling bias \citep{cai2022}. \citet{hardlessons} relabel LLM-identified false negatives in widely used training data for $0.7$--$1.4$ nDCG@10 on BEIR, \citet{tirskikh2025} filter hard negatives with an LLM, and Gecko \citep{gecko} has an LLM choose both positives and negatives among retrieved candidates, finding that the best passage often differs from the labelled one. In multimodal embedding, U-MARVEL \citep{umarvel} finds the hardest negatives unusable without threshold filtering, and MLLM judges filter mined negatives and, in one case, supply soft labels \citep{unimev2,magicmm,pixelrag}. In visual document retrieval, DSE \citep{dse} keeps a mined candidate only if the page's text lacks the answer string, a text filter we did not run as a control, and DocReRank \citep{docrerank} generates hard negative queries instead.

\paragraph{Examining LLM judgements for contrastive training.} The LLM-judge literature recommends graded labels
over binary verdicts: \citet{zhuang2024beyond} find that fine-grained relevance levels improve
zero-shot rankers over yes/no prompting, and UMBRELA \citep{umbrela} reproduces a four-point
commercial assessor with an LLM and reports agreement with human judges. Those studies ask whether a judge's labels are \emph{right}; ours become a training target, which raises a second question: what shape they have once the loss consumes them (\S\ref{sec:anatomy}).

\section{Experimental setup}
\label{sec:setup}

Within a seed every arm shares the student initialisation, the same training rows in the same
order, one optimizer schedule and one evaluation harness. An \emph{arm} is any trained
configuration, a \emph{signal} is one of N, S, A or D, and the \emph{baseline} is the InfoNCE-only
run at the same seed.

\paragraph{Student.} ColQwen2.5-3B, a late-interaction retriever in the ColPali architecture \citep{colpali} built on \texttt{Qwen2.5-VL-3B-Instruct} \citep{qwen25vl}, with MaxSim relevance $s(q,d)=\sum_i \max_j \langle \mathbf{q}_i, \mathbf{d}_j\rangle$ \citep{colbert}. We train LoRA adapters \citep{lora} on the language tower with the vision tower frozen, from the backbone rather than a released ColQwen checkpoint, so that no signal corrects a model already trained on this data.

\paragraph{Teacher.} \texttt{Qwen2.5-VL-72B-Instruct} supplies relevance scores and region descriptions. Scoring is pointwise: given the page and the query, it returns an integer rating from 0 to 100, divided by 100 (the prompt, and a yes/no readout we built but did not adopt, are in Appendix~\ref{app:teacherprompt}). Attention uses \texttt{Qwen2.5-VL-7B-Instruct}, the configuration AGREE itself selects, so the teacher is matched across the other three signals.

\paragraph{Data.} The train split of the ColPali training set \citep{colpali}: $118{,}195$ (query, page) pairs over $38{,}866$ pages, so roughly three queries share each page. Candidate pools come from \texttt{jina-embeddings-v4} \citep{jinav4}, a public 4B retriever that yields single- and multi-vector embeddings in one pass, which makes two-stage mining cheap; the result is a frozen top-100 per query, of which the teacher scores 28.

\paragraph{The four signals.} Every arm optimises $\mathcal{L}_{\text{InfoNCE}} +
\lambda\,\mathcal{L}_{\text{aux}}$, where the base term is InfoNCE over in-batch MaxSim scores \citep{infonce}, not ColPali's pairwise loss on the hardest in-batch negative \citep{colpali}. N adds documents to the InfoNCE denominator and leaves
$\mathcal{L}_{\text{aux}}$ empty; S, A and D each supply one $\mathcal{L}_{\text{aux}}$ term. N
and S are candidate-level; A and D are positive-level and act on the labelled page only.

\begin{itemize}\itemsep0pt \topsep2pt
\item[\textbf{N}] \emph{Teacher-judged hard negatives.} The four highest-ranked scored candidates
the teacher rates irrelevant (score${}<{}0.5$) are appended to that query's InfoNCE denominator,
with the lowest-scored remaining candidates filling the gap where fewer than four qualify. This carries denoised hard-negative mining \citep{rocketqa} to document images.
\item[\textbf{S}] \emph{Score distillation.} The same scores, kept soft:
$\mathcal{L}_S = \mathrm{KL}(\sigma(s_{\text{teacher}}/T_t)\,\|\, \sigma(s_{\text{student}}/T_s))$
over four scored candidates, in the direction used by cross-architecture score distillation
\citep{marginmse,colbertv2}. $\sigma$ is a softmax over those four, drawn one per quartile of the scored pool. That pool
includes the labelled page, which is therefore among the four for $12.1\%$ of queries. They enter only this term, not the InfoNCE denominator.
\item[\textbf{A}] \emph{Attention grounding} \citep{agree}. Query-conditioned teacher attention
over image tokens, pooled to the student patch grid; the loss is
$1-\cos(\mathbf{r},\bar{\mathbf{a}})$ where $r_j$ is the student's mean per-patch similarity.
\item[\textbf{D}] \emph{Description alignment} \citep{realign}. The teacher writes grounded
descriptions of the positive page; the student encodes one through its query tower, and the loss
is a KL between the query-conditioned and description-conditioned distributions over in-batch
documents.
\end{itemize}

\paragraph{Recipe.} Batch 32, learning rate $5\times10^{-5}$, one 80\,GiB H100 per arm. The primary recipe is \textbf{three epochs} ($11{,}079$ steps); every result is at that budget unless stated, and the ranking of the four
signals is unchanged at one epoch. Other settings and the determinism checks are in Appendices~\ref{app:impl} and~\ref{app:determinism}.

\paragraph{Protocol.} Every arm is compared against a baseline trained with the same seed, over three seeds for most arms and two for some; we report seed means, and Appendix~\ref{app:perseed} gives every individual run behind the ViDoRe means, the teacher-free controls and S's target variants included. The \emph{verdict} column records only whether a change keeps its sign at every seed, which an arm with no effect would still do a quarter of the time on three seeds and half the time on two, so we read it as a sanity check and rest the comparison on the size of the gaps. We do not report query-level bootstrap intervals, which cover the sampling of queries for a fixed pair of checkpoints, not the training randomness that produced them (Appendix~\ref{app:determinism}). N was run at a single configuration, its $0.5$ threshold inherited from the prompt's scale. S's temperature was changed once, from $T{=}1$ to $0.02$, after a diagnostic showed the $T{=}1$ target was nearly uniform (Appendix~\ref{app:repro}); no other value was tried. A and D were each swept over three values of their most influential hyperparameter, scored on the evaluation suites; A is reported at its published weight and D at its best swept temperature, so \textbf{D's figure is optimistic} (Appendix~\ref{app:sweeps}).

\paragraph{Human audit.} Ten annotators judged $50$ queries drawn at random from the query-complete pool (labelled page and four top-ranked candidates all scored), four pages each, on the four-point UMBRELA scale \citep{umbrela} with a \emph{cannot judge} option; a ring design gives every page two annotators and ten distinct rater pairings. Annotators never saw the teacher's score or which page carried the label. A page counts as relevant at grade ${\geq}2$, matching the teacher's cut at $0.5$; intervals are cluster-bootstrapped by query (Appendix~\ref{app:study}).

\paragraph{Evaluation.} We report nDCG@5 on four benchmarks. \textbf{ViDoRe~v1} \citep{colpali} is
in-domain --- eight of its ten test sets have their training split in our corpus --- so we treat
it as a sanity check. \textbf{ViDoRe~v2} \citep{vidorev2} and \textbf{v3} \citep{vidorev3} share
no source family with training and are our primary suites; we evaluate all their languages.
\textbf{jina~VDR} \citep{jinav4} sits between the two on domain overlap
(Appendix~\ref{app:indomain} lists which test sets are in-domain). All comparisons are
paired differences within our reimplementation of the BEIR-format evaluator, which a constant
offset cannot affect (Appendix~\ref{app:harness}).

\section{Where does the teacher help?}
\label{sec:results}

Table~\ref{tab:main} reports every signal at the three-epoch recipe (individual runs in
Appendix~\ref{app:perseed}).

\begin{table*}[t]
\centering
\small
\begin{tabular}{lrrrr rrrr l}
\toprule
& \multicolumn{4}{c}{nDCG@5} & \multicolumn{4}{c}{change vs.\ baseline} & \\
\cmidrule(lr){2-5}\cmidrule(lr){6-9}
signal & v1 & v2 & v3 & jina & v1 & v2 & v3 & jina & verdict \\
\midrule
baseline & 88.2 & 55.2 & 47.4 & 79.0 & -- & -- & -- & -- &  \\
N & 89.6 & 62.6 & 52.0 & 81.8 & $+1.4$ & $+7.4$ & $+4.5$ & $+2.8$ & \consistpos \\
S & 88.9 & 63.0 & 52.7 & 81.4 & $+0.7$ & $+7.8$ & $+5.3$ & $+2.4$ & \consistpos \\
D & 88.6 & 57.8 & 48.5 & 79.1 & $+0.4$ & $+2.6$ & $+1.1$ & $+0.1$ & \consistpos \\
A & 88.4 & 55.1 & 47.1 & 79.0 & $+0.2$ & $-0.1$ & $-0.3$ & $0.0$ & \flips \\
\bottomrule
\end{tabular}
\caption{Retrieval quality after three epochs, seed means (nDCG@5). \emph{jina} is the twenty-two English sets of jina~VDR. Changes are against a baseline trained with the same seed. N and S encode four more pages per query at $4.8\times$ the step cost; \S\ref{sec:contributes} compares N with teacher-free rules at identical training compute. \emph{verdict}: whether the v2 change keeps its sign across seeds (\consistpos{} always positive, \flips{} not consistent), a descriptive check rather than a significance test.}
\label{tab:main}
\end{table*}

\paragraph{Candidate-level signals gain more, with four extra pages per query.} Both candidate-level signals improve every suite. On ViDoRe~v2 the baseline scores $55.2$; hard negatives (N) reach $62.6$ and score distillation (S) $63.0$, gains of $7.4$ and $7.8$ points, with $4.5$ and $5.3$ on v3. Which of the two leads on v2 changes with the seed; on v3 S leads at every seed. Descriptions (D) reach $57.8$ on v2, a gain of $2.6$, and add only $0.1$ on jina~VDR against N's $2.8$; attention grounding (A) we could not get to help, its v2 change within a tenth of a point of the baseline. The margins are large relative to the seed spread: every seed's v2 gain is at least $+6.9$ for N and $+6.1$ for S, against at most $+2.9$ for D and between $-2.1$ and $+1.3$ for A (Appendix~\ref{app:perseed}).

N and S also add four mined candidates per query at $4.8\times$ the step cost, so this is not yet a clean test; \S\ref{sec:contributes} separates the teacher from the candidates for N, and \S\ref{sec:anatomy} only partly for S.

\paragraph{The ordering is stable across datasets and budgets.}
N, S and D are positive on all twelve out-of-domain datasets and A changes sign across them
(Appendix~\ref{app:tables}). The ranking of the four signals is unchanged at the one-epoch budget
(Appendix~\ref{app:perseed}).

\paragraph{Why the positive-level signals add little here.} The attention maps we extract are dominated by page structure --- about two thirds of their variation belongs to the page and one third to the query --- so AGREE's cosine objective is computed mostly over structure that does not depend on what was asked (Appendix~\ref{app:attncontent}). At ReAlign's $\tau_d{=}0.01$ the description KL is nearly one-hot, so its gradient comes from a few examples (Appendix~\ref{app:repro}); D is reported at the swept $\tau_d{=}0.05$ (Appendix~\ref{app:sweeps}). Attention on top of hard negatives leaves the result where hard negatives alone put it ($+7.1$ against $+7.5$ on v2 at matched seeds); the same combination for descriptions was not run. AGREE itself reports $+2.95$ on ViDoRe~v2 with our backbone, data, teacher, readout and weight but its own recipe (batch 128, learning rate $10^{-4}$, a hardest-negative pairwise loss; single runs), which we did not run (Appendix~\ref{app:attncontent}). This describes the methods as adapted here.

\section{Is it the teacher, or the extra candidates?}
\label{sec:contributes}

N combines two things: it trains against mined candidates, and it asks the teacher which of them to keep. Five teacher-free controls keep the first and drop the second. Three take four non-gold negatives evenly spaced across a fixed window of retriever ranks (throughout, ranks count positions in the mined pool with the row's labelled page excluded, though not pages labelled for the same query string in other rows). Two apply TopK-PercPos \citep{nvretriever}, which keeps a candidate only if the \emph{miner's} score falls below $\alpha$ times the positive's, so the cut moves with the query; a threshold of this kind is what current leading visual document retrievers use \citep{colembedv2,qwen3vlemb,kovre}. We run it at its authors' $\alpha$ and at the $\alpha{=}0.864$ that matches N's rejection rate --- $41.9\%$ of the teacher-scored candidates, on the queries whose labelled page the miner retrieved, so both rules share a denominator. All controls append four candidates, matching N's exposure and compute.

\begin{table}[t]
\centering
\small
\setlength{\tabcolsep}{4pt}
\begin{tabular}{ll r rrr}
\toprule
& & rated & \multicolumn{3}{c}{change vs.\ baseline} \\
\cmidrule(lr){4-6}
arm & negatives from & relevant & v1 & v2 & v3 \\
\midrule
N-mined & ranks 1--4 & 62\% & $+0.1$ & $-0.2$ & $-2.1$ \\
N-rank & ranks 4--20 & 47\% & $+0.5$ & $+0.4$ & $-0.9$ \\
N-band & ranks 21--100 & 36\% & $+1.0$ & $+2.4$ & $+2.0$ \\
N-perc & miner $<0.95\,s(q,d^{+})$ & 45\% & $+1.8$ & $+3.3$ & $+2.1$ \\
N-perc & miner $<0.864\,s(q,d^{+})$ & 32\% & $+1.3$ & $+3.3$ & $+2.8$ \\
\textbf{N} & \textbf{teacher-judged} & 20\% & $+1.4$ & $+7.4$ & $+4.5$ \\
\bottomrule
\end{tabular}
\caption{Teacher-free negative selection at N's compute (three epochs). The first three rows take four candidates evenly spaced in a fixed rank window; N-perc applies TopK-PercPos \citep{nvretriever} at two values of $\alpha$. \emph{relevant} is the share of the negatives each rule trains on that the teacher rates ${\geq}0.5$, back-fill included; the first two windows share rank~4. Seed-mean changes against a seed-matched baseline (nDCG@5); two seeds except N and $\alpha{=}0.95$, which have three.}
\label{tab:contributes}
\end{table}

\paragraph{The hardest negatives are the worst ones.} Taking the four highest-ranked non-gold
candidates --- what hard-negative mining usually means --- is the weakest control: no better than its baseline on ViDoRe~v2 ($-0.2$, its two seeds of opposite sign) and below it on v3 ($-2.1$), as in text retrieval \citep{simans}. The teacher rates \textbf{62\%} of those candidates relevant: a retriever's top candidates are where its relevant pages are, and pushing them away unlearns them.

\paragraph{Moving off the top helps, and moving further helps more.} Four candidates spread over ranks 4--20, where the teacher's filter typically lands, recover roughly the baseline ($+0.4$ on v2, $-0.9$ on v3) with $47\%$ rated relevant; ranks 21--100 do better ($+2.4$ and $+2.0$) with $36\%$. The false-negative rate falls as retrieval improves across the three windows, but deeper windows are also wider (4, 17 and 80 ranks) and so draw less similar pages, and these runs cannot separate the two effects.

\paragraph{Adapting the cut per query helps, and still does not reach the teacher.} The adaptive rule is the strongest control: $+3.3$ on v2 against the best window's $+2.4$, while admitting \emph{more} teacher-rated-relevant negatives ($45\%$ against $36\%$), and its two settings score alike despite a $13$-point gap in that rate. So the false-negative rate does not by itself account for these results: moving the cut per query buys something a single window, applied to every query alike, cannot.

\paragraph{The teacher beats the best rule that does not use one.} $7.4$ against $3.3$ on ViDoRe~v2 at a matched rejection rate, and $4.5$ against $2.8$ on v3; N leads every teacher-free rule at every seed on both suites (Table~\ref{tab:perseedarms}). The two also disagree about which candidates are safe: rate-matched, their four negatives are identical for only \textbf{$15\%$} of queries and overlap on $1.3$ of four, so the gap is not one of degree within a shared ordering. Since the strongest control is the rule used in practice, the margin is the value of consulting a judge rather than the retriever's own scores.

So the teacher's contribution to N is a per-query choice of negatives that no fixed or adaptive rule we tried matches: with the candidate pool and compute fixed, it is worth \textbf{$+4.1$} on ViDoRe~v2 (per seed against $\alpha{=}0.95$: $+5.6$, $+2.4$, $+4.2$) and \textbf{$+1.7$} on v3 (against a two-seed control; $+2.4$ at three matched seeds). Descriptions gain $+2.6$ and $+1.1$ and attention nothing measurable, but from the InfoNCE baseline, not on top of mined negatives, and D was never combined with them; the teacher's increment exceeds D's gain on average but not at every seed.

\section{How incomplete are the training labels?}
\label{sec:labels}

This section measures how much there is for the teacher's per-query decision to do on this corpus: how its queries were written, what the teacher rates relevant, and what annotators judge relevant.

\paragraph{The queries were written against one page, not against a collection.} The ColPali
training set is assembled from document-VQA datasets and synthetic collections, whose questions
were each written or generated against a single
page (\S\ref{sec:intro}). VisRAG already filters such context-dependent queries with an LLM \citep{visrag}. The effect is visible in the corpus itself: \textbf{2{,}736} query strings recur, $2{,}620$ against more than one page and one against $70$; $6{,}377$ rows repeat a string.

\paragraph{The teacher rates many pages relevant per query, and how many depends on the source.}
Over the pool it scores, the teacher rates a median of \textbf{10} non-gold pages relevant per
query, and only \textbf{12.5\%} of queries have none. The retriever's top non-gold candidate ---
the first page any hard-negative miner would take --- is rated relevant by the teacher for \textbf{70.1\%} of queries (RocketQA's hand check found about $70\%$ of such MS~MARCO passages relevant; \citealp{rocketqa}), and the rate tracks how the questions were written: $83\%$ for TAT-DQA, $81\%$ DocVQA,
$65\%$ synthetic PDF, $62\%$ arXiv-QA and $43\%$ Infographic-VQA, whose questions name specific
visual content (Appendix~\ref{app:falseneg}).

\paragraph{Annotators agree, on the pages they were shown.} The teacher's leniency is not the explanation. On the $29$ audited queries that show a complete top four, annotators marked \textbf{$2.04$} of those four relevant on average ($[1.46, 2.61]$, cluster-bootstrapped by query), and the dataset labels none of them, so a miner that takes the top four trains against all of them. Our pre-specified test was the conservative one, whether the mean exceeds a single relevant page rather than zero, and the interval clears it.

\paragraph{The filter cannot remove them all.} On \textbf{55.7\%} of queries at least three of the four top-ranked candidates are rated relevant, and for $45\%$ of those --- about a quarter of all queries --- the $28$ scored candidates do not contain four the teacher calls irrelevant, so N back-fills with pages it rated relevant itself. Hence \textbf{20\%} of N's negatives are still teacher-rated relevant (Table~\ref{tab:contributes}; by stratum, and an inconclusive attempt to localise N's gain, in Appendices~\ref{app:strata} and~\ref{app:localisation}).

\paragraph{These are false negatives, not vague queries.} ``What is the date?'' may sound vague, but whether a given page answers it is checkable; it is simply answered by many pages. Much of the contamination measured here looks like this: the rejected candidates answer the query as posed, so training against them teaches the retriever that a relevant page is wrong. Whether a query has one determinate answer is a separate property, and \S\ref{sec:validation} shows it governs how far the teacher can be trusted.

\section{What reaches the student?}
\label{sec:anatomy}

If what the teacher contributes is a missing relevant/irrelevant label, then S's graded targets
should add little beyond that split. This section checks that for the rating prompt used here.

\paragraph{Under a greedily decoded rating the signal is near-binary.} Asked for a $0$--$100$ relevance rating, decoded greedily, the teacher puts $51.7\%$ of its $3.3$M scores at exactly $0$ and $29.9\%$ at exactly $1$. This is a property of the readout, not of
the teacher: the same model asked for a yes/no logprob returns a continuous score, which ties on
$0.01\%$ of gold-versus-mined comparisons against the rating's $30.7\%$ and separates the labelled positive from its mined candidates better (Appendix~\ref{app:teacherprompt}).

\paragraph{Over the four candidates S distils, the target is often constant.} S distils a softmax
over $K{=}4$ candidates, one per quartile of the scored pool. Because the ratings are nearly
binary, the target vector is \textbf{constant for $42.6\%$ of queries} --- $27.0\%$ all-zero,
$15.1\%$ all-one and $0.5\%$ at another value, most often $0.95$. A constant vector is uniform
after a softmax at any temperature, so on those queries the KL term says nothing about which
candidate is more relevant and only pulls the student towards spreading its mass equally over the four. Where it is not constant, $T{=}0.02$ brings it close to an argmax with ties: a $0.1$ rating gap is a mass ratio of about $150$.

\begin{table}[t]
\centering
\small
\begin{tabular}{lrrrc}
\toprule
target & v1 & v2 & v3 & seeds \\
\midrule
S \ (teacher ratings) & $+0.74$ & $+7.80$ & $+5.31$ & 3 \\
\quad flat target, no teacher & $-0.63$ & $-3.08$ & $-6.00$ & 2 \\
\quad S-masked: constant rows dropped & $-0.42$ & $+10.50$ & $+6.50$ & 3 \\
\quad partition only, gradation dropped & $+0.88$ & $+5.58$ & $+5.63$ & 2 \\
\quad\emph{S again, at the partition arm's two seeds} & $+0.91$ & $+7.33$ & $+5.03$ & 2 \\
\bottomrule
\end{tabular}
\caption{S's supervision, partitioned by target. All rows share the loss, its weight $\lambda_s{=}1$ and the candidate forward passes; only the target differs, though S-masked also lowers the effective weight (Appendix~\ref{app:repro}). Changes against a seed-matched baseline (nDCG@5). The last row repeats S at the two seeds the partition-only arm ran, the comparison \S\ref{sec:anatomy} quotes.}
\label{tab:spartition}
\end{table}

\paragraph{A constant target is a uniformity penalty, not merely uninformative.} Two arms bound what those rows cost; both keep S's loss, weight, candidates and forward passes and change only the target (Table~\ref{tab:spartition}). A flat target on every row, with no teacher at all, loses $3.1$ points on ViDoRe~v2 and $6.0$ on v3 at both seeds. That arm is S's teacher-free control, so S's gain is not compute and not candidate exposure; but it is a weaker control than the rules of \S\ref{sec:contributes} are for N, since a flat target carries no information at all, and how much an informative teacher-free target such as the miner's own scores would recover is untested. Zeroing the KL on exactly the constant rows instead --- \emph{S-masked} --- \textbf{beats S by $2.7$ points} on v2 ($+10.50$ against $+7.80$, higher at every seed) and by $1.2$ on v3: rows that carry no relevance information are not inert, and the effective-weight change masking also causes does not account for it (Appendix~\ref{app:repro}). The arm was found post hoc and scored on the evaluation suites, so its size is exploratory, though its sign holds at every seed.

\paragraph{Most of the gain survives without the gradation.} Replacing the graded scores on the informative rows with the partition alone --- mass $1/|R|$ on the teacher-relevant set $R$, uniform when $R$ is empty or holds all four --- keeps most of what S gains: at the two seeds both arms share, $76\%$ of S's v2 gain ($+5.58$ against $+7.33$), nearly all of it on v1 ($+0.88$ against $+0.91$) and all of it on v3 ($+5.63$ against $+5.03$). On v2 the partition trails S at both shared seeds (by $1.2$ and $2.3$), so the gradation may add a little; two seeds cannot size it, nor does the comparison isolate it: the partition also makes uniform every row whose four candidates fall on one side of the threshold ($54.3\%$ of rows, against $42.6\%$ for S), so part of the shortfall may be the uniformity penalty above; the arm is post hoc.

\paragraph{N and S use one partition in two ways.} The softmax also removes the absolute level:
$[0,0,0,0]$ and $[1,1,1,1]$ give the same target, so S cannot tell a query whose four candidates
are all irrelevant from one whose four are all relevant, while N, which thresholds the score, can.
The two consume the same incomplete labels in opposite directions, the elimination-versus-attraction choice studied for false negatives in contrastive learning \citep{huynh2022} --- N keeps teacher-rated
relevant pages out of the negatives, S raises their target (for $51.4\%$ of queries its four
candidates include a non-gold page scored ${\geq}0.9$) --- consistent with two such different losses landing close together.

\section{When can the teacher be trusted?}
\label{sec:validation}

The audit of \S\ref{sec:setup} tests whether the teacher's judgements are right (Appendix~\ref{app:study}).

\begin{table}[t]
\centering
\small
\begin{tabular}{lrr}
\toprule
& value & $n$ \\
\midrule
precision of protection & 71\% [59, 80] & 197 \\
\quad on \emph{specific} queries & 83\% [71, 92] & 138 \\
\quad on \emph{broad} queries & 42\% [24, 58] & 59 \\
\addlinespace[2pt]
miss rate & 14\% [8, 22] & 141 \\
\midrule
teacher--human $\kappa$ & 0.54 & 338 \\
human--human $\kappa$ (mean of 10 pairings) & 0.41 & 161 \\
\quad range across pairings & $-0.13$ to $0.87$ & \\
\midrule
relevant pages per query, of four & 2.04 & \\
\bottomrule
\end{tabular}
\caption{Human audit: ten annotators, 50 queries, 400 judgements, every page judged by two annotators. \emph{Precision of protection} is the share of the candidates the teacher protects (score ${\geq}0.5$) that an annotator calls relevant (grade ${\geq}2$), split by query type in the two rows beneath it; \emph{miss rate} is the share of the audited top-four candidates it rejects that an annotator judges relevant. Intervals are $95\%$ cluster-bootstrapped by query; the \emph{specific}/\emph{broad} split was fixed before any grade was read. Both $\kappa$ are over mined candidates only, with labelled positives excluded: teacher--human over every such judgement, human--human over the pages each ring pair shares.}
\label{tab:study}
\end{table}

\paragraph{The filter is right about most of what it protects.} Of the candidates the teacher rates relevant --- the ones N declines to train against --- $71\%$ are judged relevant by an annotator too ($[59,80]$, Table~\ref{tab:study}). Conversely, $14\%$ of the audited candidates it rates irrelevant --- the pages N trains against among the top four --- are judged relevant: the teacher's own misses. These are separate from the back-fill of \S\ref{sec:labels}, pages the teacher itself rates relevant ($20\%$ of N's negatives), which the audit did not sample.

\paragraph{The teacher sits inside the range of human agreement.} Mean human--human $\kappa$ over the ten rater pairings is $0.41$ and teacher--human $\kappa$ $0.54$; these are different estimators (a mean of ten pairwise $\kappa$ over $161$ doubly graded pages, one pooled over $338$ judgements) and $\kappa$ moves with prevalence, so we read them descriptively. Per-pairing $\kappa$ runs from $-0.13$ to $0.87$, and leaving out each annotator keeps the teacher above the human mean, by $+0.05$ to $+0.25$. Annotators' relevant-rates run from $11\%$ to $81\%$, mixing strictness with the block of ten queries each drew. The weaker claim we need holds: the teacher's disagreements with annotators are not of a different order from theirs with each other.

\paragraph{The teacher is reliable where the question has one determinate answer.} Before any
grade was read, each query was classified as \emph{specific} or \emph{broad}:
whether it admits a single determinate answer or several different ones. On the $33$
specific queries the teacher's protected candidates are $83\%$ relevant $[71,92]$; on the $17$
broad ones, $42\%$ $[24,58]$, near the base rate of relevance among these candidates, and the
intervals do not overlap. Whether a page
answers ``what was the valuation allowance decrease in 2019'' is checkable; whether it answers
``how is accuracy of the data ensured'' is not.

\paragraph{The threshold trades precision against misses.} Precision of protection rises with the cut --- $71\%$ at ${\geq}0.5$, $80\%$ just above it, $85\%$ at $1.0$ --- mostly because the hedge value $0.5$ is only $28\%$ relevant ($36$ audited pairs); but every page a higher cut releases becomes an eligible negative; we did not retrain N at other cuts, and S never thresholds.

\paragraph{The hedge falls mostly on broad queries.} Of the queries where the teacher hedges, $69\%$ are broad, against $22\%$ of those where it does not (Fisher exact $p{=}0.005$; a post-hoc analysis that reads no human grades and whose two inputs were fixed before any page was judged, Appendix~\ref{app:study}). This is consistent with \S\ref{sec:anatomy}'s near-binary readout: the teacher treats the question as yes/no and hedges where it has no single answer.

\section{Implications and scope}
\label{sec:conclusion}

\paragraph{What follows for practice.} On corpora like this one, a judge over the mined candidates beat every teacher-free rule we tried on both primary suites, though whether enriching the positive adds on top of mined negatives is untested; N and S cost one $46$ GPU-hour annotation pass and $4.8\times$ the training step, the most expensive single run of the four (Appendix~\ref{app:cost}). Without a judge, do not train on the top-ranked candidates: they are the worst negatives we tested, no better than the baseline on v2 and below it on v3. When distilling a judge's scores, check what the constant-target rows are doing: masking them in one post-hoc arm gained $2.7$ on v2 and $1.2$ on v3 but cost $1.2$ in domain (Table~\ref{tab:spartition}).

\paragraph{Threats to validity.} One student, one corpus, one late-interaction architecture and one teacher (7B for attention, 72B otherwise); a larger student or a single-vector retriever could reorder the signals, most plausibly for descriptions, whose published form is single-vector. Our recipe also leaves the baseline below the released colqwen2.5-v0.2 (about $61$ on v2 as reported by \citealp{vidorev2}; Appendix~\ref{app:harness}), whose recipe we did not reproduce, so effects are measured from a weaker start, and mined negatives may matter less at larger batches. D's temperature was tuned on the evaluation suites, and A and D are evaluated through page retrieval, not the grounding one of them targets. S's flat-target control rules out compute alone, but no informative teacher-free target was tried, only attention was combined with mined negatives, and no text-only judge or answer-string filter was compared. The coarseness of \S\ref{sec:anatomy} belongs to the rating prompt and temperature. Two to three seeds give a consistent sign and ordering of the mean gains --- for N and S several times the spread between seeds --- but not precise effect sizes, and the audit covers only top-ranked candidates (Appendix~\ref{app:limitations}).

\paragraph{Conclusion.} What transferred from the teacher here was a yes/no judgement on the unlabelled pages a miner retrieves: annotators find about two of its top four relevant, and a judge that says which ones to train against beats every teacher-free rule we tried on both primary suites. Under our rating prompt and temperature, what reaches the student is close to that bare relevant/irrelevant split. We expect that advantage to shrink on better-labelled corpora. The lesson we take is about labels as much as teachers: however rich the signal around the labelled page, check what the negatives are teaching.

\section*{Use of AI}
An AI coding assistant was used throughout this work: for implementing and testing the four
training signals and the evaluation harness, for the analysis scripts that produce every table,
for proposing and refining controls, for interpreting results, and for drafting and revising the
text of this paper. It was also used to review drafts and to propose the restructuring the
current version follows. Two uses belong to the method rather than the writing: the teacher's
relevance scores and region descriptions are generated by a vision--language model
(\S\ref{sec:setup}), and the page transcripts in the annotation interface are machine-generated
OCR (Appendix~\ref{app:study}). No AI produced any of the $400$ human judgements or the
\emph{specific}/\emph{broad} classification of queries, and every number reported here is emitted
from the evaluation result files rather than written by a model. The analysis of the human study
was pre-specified and frozen before any judgement was read, and its one post-hoc analysis is
labelled as such (\S\ref{sec:validation}). All AI-assisted text, code and analysis were reviewed
by the authors, who take full responsibility for the content of this paper, including every number
reported in it.

\section*{Reproducibility}
\label{sec:repro}
The following are released: the training and evaluation code; the teacher's relevance scores
for all $3.3$M scored (query, candidate) pairs and its region descriptions for all $118{,}195$
training rows; the frozen mined candidate pools; the human study's protocol and its $400$
judgements; and trained adapters, at one seed each, for the baseline, both candidate-level signals
(N and S), the teacher-free TopK-PercPos control at $\alpha{=}0.95$, and the S-masked arm of
\S\ref{sec:anatomy}.
The code, the data and the human study are at \url{https://github.com/elastic/vdr-teacher-signals},
which regenerates every table in this paper from the included run outputs; the trained
adapters are linked from the same repository.
Together these let the judged negatives and the score targets be reproduced, or reused for a
different student, without access to a 72B teacher, and let the reported numbers for those arms be
checked at one seed without retraining. We release these five adapters rather than every arm we trained, to keep the
release usable. Within this paper: Appendix~\ref{app:teacherprompt} gives the teacher prompt, the
parsing rule and the alternative yes/no readout we built but did not adopt; Appendix~\ref{app:harness}
gives the evaluation harness and its offset against published numbers;
Appendix~\ref{app:determinism} shows that two runs of an identical configuration match on every
quantity we monitor, and reports the control behind our reading of query-level bootstrap intervals;
Appendix~\ref{app:perseed} lists the individual runs behind every seed mean the main text reports,
the teacher-free controls and S's target variants included;
Appendix~\ref{app:cost} gives the full cost breakdown including teacher token counts;
Appendix~\ref{app:study} gives the annotation rubric, the ring assignment and the per-annotator
breakdown; and Appendix~\ref{app:repro} records three defects we found in our own implementations,
each of which produced a plausible result before it was caught.

\section*{Ethics}
The human study collected relevance judgements over pages of publicly available documents drawn
from an existing, publicly released training corpus. The ten annotators were unpaid volunteers
rather than crowdworkers: one is an author of this paper, five are colleagues and four are
personal acquaintances. All consented to the use and pseudonymised release of their judgements,
and the nine who are not authors also to being named in the acknowledgements; the task took
about twenty-five minutes. No annotator saw the teacher's score or which page
carried the dataset label, and dropping any single annotator leaves relevant-pages-per-query
within $1.93$--$2.24$ (Appendix~\ref{app:study}). No personal data was collected from annotators, who are identified in
the released judgements only by pseudonymous identifiers (\texttt{v01}--\texttt{v10}); no
demographic information was recorded. Annotators judged query-page pairs; they were not asked to produce or evaluate generated content, and the machine transcript shown beside each page served only as a search aid (Appendix~\ref{app:study}). The released judgements contain no free
text. We report the full distribution of annotator behaviour, including the two annotators whose
relevant-rates sit at the extremes of the range, rather than excluding outliers
(Appendix~\ref{app:study}).

\section*{Acknowledgements}
We thank Mohammad Kalim Akram, David Eristavi, Quentin Herreros, Florian Hoenicke,
Vakhtang Kotoreishvili, Andreas Koukounas, Dachi Kurtskhalia, Georgios Mastrapas and
Marta Rozzoni, who volunteered the relevance judgements of \S\ref{sec:validation}, and
Scott Martens for his assistance with the writing.

\bibliographystyle{iclr2027_conference}
\bibliography{refs}

\appendix
\section{Three implementation pitfalls in late-interaction training}
\label{app:repro}

All three were invisible in the loss curve and were caught by inspecting a distribution rather
than a scalar. We record them because we expect them to be common.

\paragraph{A reduced-precision output layer that stopped learning.} Two runs of a byte-identical
configuration finished 2.31 nDCG apart, which we first read as benchmark noise. We adapt the
language tower with LoRA, which the adapter library upcasts to fp32, but train the output
projection through \texttt{modules\_to\_save}, which copies the module at the base model's load
dtype, bf16. One Adam step at $\text{lr}{=}5\times10^{-5}$ is of order $5\times10^{-5}$; the layer
initialises uniformly on ${\pm}0.022$, where one bf16 unit in the last place runs from
$6\times10^{-5}$ to $1.2\times10^{-4}$, so a step of that size falls below half a unit for the
larger weights and rounds away. After 3{,}693 steps, \textbf{43.6\% of the projection's weights
were still bit-identical to their initialisation}; after the fix, 0.00\% within 20 steps. Because
bit-level nondeterminism decided \emph{which} half received a representable update, two identical
configurations produced models 16\% apart in relative $L_2$ on that layer while their losses
agreed to ${\approx}0.03$. Fix: fp32 master weights for trainable projection modules. The
reference ColPali implementation avoids this by including the projection in the LoRA target regex.
Worth reporting after any short fine-tune: the fraction of each trainable module's weights that
differ from initialisation.

\paragraph{S: a distillation target that was almost uniform.} S's KL compares two softmax
distributions over $K{=}4$ documents. At the temperature we first used ($T_t{=}1$) the teacher's
target has mean entropy 1.349 against 1.386 for uniform --- \textbf{97.3\% of the entropy of pure
noise} (Table~\ref{tab:temperature}). The arm was regularising the student towards uniformity over
its own candidate pool. At $T_s{=}T_t{=}0.02$, the temperature InfoNCE already uses, the target
falls to 59.1\% of uniform entropy and the optimised loss rises $11\times$ (mean distillation loss $0.053 \to 0.572$). Both entropies are measured over the pool the collator
actually builds --- one candidate drawn per quartile of the scored pool --- rather than the four
highest-ranked candidates; the latter understates the spread of the target and so overstates how
much $T{=}0.02$ sharpens it. The correction changes the paper's answer: S goes from sign-flipping
to indistinguishable from N.

\paragraph{Masking constant rows also lowers S's effective weight.} The KL is averaged over every row, so zeroing it on the $42.6\%$ that are constant leaves the surviving rows carrying $0.574$ of S's weight. To separate that from the masking itself we ran S at $\lambda_s{=}0.576$, essentially that weight, with the teacher target kept on all four candidates: at the shared seed it moves S by $-0.13$ on ViDoRe~v2 ($+5.98$ against $+6.11$), where masking moves it by $+4.57$. The weight change is therefore not the explanation, though that control ran at one seed only.

\paragraph{D: a distillation target that was almost one-hot.} At the inherited $\tau_d{=}0.01$ both
sides of D's KL are effectively one-hot ($2.4\%$ and $3.3\%$ of uniform entropy), and in a
converged checkpoint the query and the description select the same page $96.3\%$ of the time. The
per-example loss is therefore not small but \emph{sparse}: median $0.0004$, $76\%$ of examples
below $0.01$, and the worst $4\%$ carrying $67\%$ of the term. The batch-mean was a healthy
$0.178$ throughout --- a mean cannot distinguish a small loss everywhere from no loss almost
everywhere and a spike on a few. At $\tau_d{=}0.05$ the worst $4\%$ carry $9.3\%$, i.e.\ little
concentration; that is the setting we report (Appendix~\ref{app:sweeps}).

We can establish that in our configuration the term acts on few examples, not why $0.01$ was
reported
to work. \citet{realign} apply one temperature to both their contrastive and KL terms where we set
the KL to their $0.01$ with our contrastive term at $0.02$, and their student is single-vector
where ours is late-interaction. D's result at ReAlign's temperature is therefore evidence against description
alignment at this temperature in this architecture, not against the method.

The general hazard: a distillation temperature is a property of the loss \emph{together with the
scores it is applied to}, so it does not transfer with the equation.

Table~\ref{tab:temperature} gives the entropy of S's target distribution against the
distillation temperature, which is the diagnostic that prompted the change from $T{=}1$
to $0.02$.

\begin{table}[H]
\centering
\small
\begin{tabular}{rrrr}
\toprule
$T$ & entropy & max prob. & \% of uniform \\
\midrule
\bfseries 1 & \bfseries 1.349 & \bfseries 0.314 & \bfseries 97.3\% \\
0.5 & 1.255 & 0.377 & 90.5\% \\
0.1 & 0.908 & 0.506 & 65.5\% \\
0.05 & 0.866 & 0.524 & 62.4\% \\
\bfseries 0.02 & \bfseries 0.819 & \bfseries 0.542 & \bfseries 59.1\% \\
\midrule
uniform & 1.386 & 0.250 & 100\% \\
\bottomrule
\end{tabular}
\caption{Entropy of the teacher's target distribution over the $K{=}4$ scored pool as a function of the distillation temperature. The pool is the rank-stratified draw the collator actually makes --- one candidate per quartile of the scored pool --- not the top four by retriever rank. At the temperature our score distillation arm originally used ($T{=}1$) the target retains 97.3\% of the entropy of a uniform distribution, i.e.\ it is almost information-free; at $T{=}0.02$, which matches the InfoNCE temperature, it falls to 59.1\%.}
\label{tab:temperature}
\end{table}

\section{Teacher relevance scoring}
\label{app:teacherprompt}

The page is placed before the query so that vLLM reuses the prefix cache when one page is scored
against several queries, which it usually is --- roughly three queries share each page. The prompt
is:

\begin{quote}\small\ttfamily
You are judging visual document retrieval relevance. Above is a document page.\\
Query: \{query\}\\
Rate how relevant this page is to the query on a scale from 0 (completely irrelevant) to 100
(directly contains the answer). Respond with only the integer.
\end{quote}

We decode at temperature $0$ and divide the returned integer by 100. A score is therefore a
normalised rating in $[0,1]$, not a calibrated probability: we do not check that pages scored
$0.7$ are relevant $70\%$ of the time, and nothing here establishes that they are.

We also built a yes/no readout, taking
$P(\text{yes})=e^{\ell_{\text{yes}}}/(e^{\ell_{\text{yes}}}+e^{\ell_{\text{no}}})$ from the
first-token logprobs, and originally rejected it on a positive-in-top-3 rate: the labelled
positive ranked in the teacher's own top three for $74.4\%$ of queries under that readout against
$96.4\%$ for the rating. \textbf{That comparison is invalid:} The rating is
coarsely quantised, so the gold page ties one or more mined candidates at the identical integer in
\textbf{$30.7\%$} of gold-versus-mined comparisons, and the metric breaks those ties in gold's
favour by construction; continuous logprobs tie in $0.01\%$ of comparisons and so receive none of
that credit. Breaking ties the other way instead drops the rating to $53.2\%$ and leaves the
logprob readout unchanged at $74.4\%$ --- the ordering is an artefact of the tie rule, not a
property of the readouts.

On a metric that handles ties explicitly --- per-query tie-aware AUC of gold against its mined
candidates, i.e.\ Mann--Whitney $U$ with half credit for ties --- the \emph{logprob} readout is
the better discriminator: $0.880$ against $0.825$ for the rating, a difference of $+0.055$ with a
$95\%$ query-bootstrap interval of $[+0.040,+0.071]$ on the same $500$ queries and $9{,}577$
pairs. We therefore do not claim the rating is the better readout. We keep it for three reasons
that do not depend on that claim. First, both signals need the \emph{magnitude} and not merely an
ordering --- N thresholds it and S distils a distribution over it --- and graded prompts are what the LLM-judge literature recommends for exactly this purpose \citep{zhuang2024beyond,umbrela}, although \citet{zhuang2024beyond} read the grade from the probabilities of the label tokens rather than by greedy decoding, as we do.
Second, re-annotating $3.31$M pairs and retraining N and S under the other readout costs on the
order of $700$ GPU-hours, which we judged better spent on the controls in \S\ref{sec:contributes}.
Third, only N and S consume the rating, and both consume the same one, so the readout cannot be a
confound between them. Whether the other readout would change their margin over A and D we cannot
say: discriminating labelled positives better need not make a better training target. We keep the
rating for the experiments reported here, and performance under the alternative readout is untested.

Replies with no integer, or an integer above 100, are treated as unparseable and dropped rather
than clipped: a single bogus value would dominate any softmax S builds over its pool. This affects
\textbf{0.64\%} of the 3.31M scored pairs. Pages are capped at ${\approx}768$ visual tokens.

\section{Which test sets are in-domain}
\label{app:indomain}

\textbf{ViDoRe~v1.} Eight of ten test sets have their training split in the ColPali corpus. Four
are visible in the source names (\texttt{docvqa}, \texttt{tatdqa}, \texttt{Infographic-VQA},
\texttt{arxiv\_qa}); the other four are not --- the corpus's largest source, labelled \texttt{pdf}
(45{,}718 rows), holds the training splits of the four \texttt{syntheticDocQA} sets v1 tests on.
Only \texttt{shiftproject} and \texttt{tabfquad} have no counterpart. \textbf{v2 and v3} share no
source family with training. \textbf{jina~VDR}: eight of the twenty-two English sets are revised
v1 tasks (\texttt{arxivqa}, \texttt{docvqa}, \texttt{infovqa}, \texttt{tatqa}, the four
\texttt{docqa\_*}) and fourteen are new. We report the suite as a whole rather
than split by that contrast.

\section{Harness calibration}
\label{app:harness}

We reimplement the ViDoRe BEIR-format evaluator rather than calling it, so absolute scores are not
automatically comparable to published ones. We did not evaluate the released \texttt{colqwen2.5-v0.2} on v2 or v3 in our harness; the ViDoRe~v2 paper reports it at $68.4$, $61.1$, $57.4$ and $56.5$ on the four datasets we evaluate, $60.9$ on average \citep{vidorev2}, $5.7$ points above our baseline. Our shared recipe was sized so that every arm fits one device at batch 32, not tuned for absolute scores. On ViDoRe~v1 the released \texttt{colqwen2.5-v0.2} checkpoint scores $88.5$ here, about $0.9$ nDCG below its published
figures. On v2 we can estimate the offset directly: the strongest publicly reported retriever at 4B
parameters or below on the ViDoRe~v2 leaderboard as of September 2026,
\texttt{tomoro-colqwen3-embed-4b} \citep{tomorocolqwen3}, scores $64.50$ under our harness against
$64.69$ on that leaderboard, a difference of $0.19$. That is one model, so it bounds the offset
only approximately. Every comparison in this paper is a
paired difference computed \emph{within} our harness against a seed-matched baseline, so a
constant offset cannot affect any of them; the calibration matters only when reading an absolute
number next to a published one.

\section{Implementation details}

The train split lists $39{,}239$ distinct \texttt{image\_filename}s, which the miner deduplicates by image content hash to the $38{,}866$ pages of \S\ref{sec:setup}; $117$ of its rows repeat a (query, page) pair exactly.
\label{app:impl}

Details omitted from \S\ref{sec:setup}; none of them varies across arms.

\textbf{Student.} LoRA \citep{lora} at rank $32$, scaling $32$, dropout $0.1$, on the language
tower only; the vision tower is frozen and the freshly initialised 128-dimensional output
projection is trained fully rather than through an adapter. Together that is 60.1M trainable
parameters of 3.8B.

\textbf{Teacher.} The 72B teacher runs AWQ 4-bit. Replies that do not parse as an integer rating
are dropped, which is $0.6\%$ of pairs; the affected query is then scored on fewer candidates
rather than on a guessed value.

\textbf{Data.} Of the $118{,}195$ (query, page) pairs, a frozen manifest of $118{,}184$ rows is
shared by every arm --- the eleven dropped rows have no usable page. Teacher scoring covers the
full set, and $118{,}194$ queries end up with at least one scored non-gold candidate.

\textbf{Recipe.} Linear schedule with $2.5\%$ warmup, bf16 with gradient checkpointing. Three
epochs is $11{,}079$ optimizer steps and one epoch $3{,}693$. The in-batch contrastive loss runs
at temperature $0.02$, and S uses $T_s{=}T_t{=}0.02$; Appendix~\ref{app:repro} reports the target
entropy that setting produces, which is how the temperature was chosen. Batch 64 is infeasible for
the candidate-level arms: with four extra pages per query they already occupy 75.8\,GiB of an
80\,GiB device.

\textbf{Loss weights.} $\lambda_s{=}1$ for score distillation, $\lambda_a{=}0.1$ for attention
grounding and $\lambda_d{=}0.2$ for description alignment, each the value its source reports as
best. D is reported at the swept $\tau_d{=}0.05$ rather than ReAlign's $\tau_d{=}0.01$
(Appendix~\ref{app:sweeps}).

\section{Determinism}
\label{app:determinism}

With the dtype fix above, \texttt{CUBLAS\_WORKSPACE\_CONFIG=:4096:8}, deterministic kernels and
RNG state restored across resumption, two runs of an identical configuration reproduce each other
exactly on everything we monitor --- verified by training the baseline twice on \emph{different
machines} and obtaining identical loss and gradient norm at step 50, to every printed digit, and
identical evaluation scores. We did not hash the weights, so this is reproducibility of the
observed quantities rather than a proof of bit-level identity. An arm-versus-baseline delta is
therefore attributable to the arm's change. It is not thereby robust to reseeding, which is a
separate question (\S\ref{sec:setup}), and a query-level paired bootstrap is easy to mistake for an
answer to it. Such an interval covers the sampling of queries \emph{conditional on} the two
checkpoints being compared; it says nothing about the training randomness that produced them.
Feeding a differently-seeded baseline into it \emph{as though it were a treatment} separates the one-epoch baselines at seeds $0$ and $1$ by $4.5$ points on ViDoRe~v2 and $2.6$ on v3 with queries pooled across each suite's datasets, both intervals clear of zero; averaging per dataset instead, as Table~\ref{tab:perseed} does, gives $2.7$ and $2.6$. On the in-domain v1 the pooled gap is $0.1$ and its interval contains zero. The bootstrap is not at fault there: the two checkpoints
really are different functions, and the gap between them on these queries really is that large. What
is missing is anything to attribute the gap to, since no intervention was applied. This is why we
compare runs that share a seed and read a query-level interval as a statement about the queries
rather than about the recipe.

\section{What an attention map contains}
\label{app:attncontent}

Most of a map is the same whatever the query is. Roughly three queries share each page, so maps
for one page under different queries can be compared: mean cosine $0.689$, against $0.330$ for
different pages. Decomposing the variance of normalised maps over shared pages attributes about
\textbf{two thirds to the page and one third to the query} ($65\%/35\%$).\footnote{The naive estimate is $77\%/23\%$ but is biased: with $k$ maps per page the page mean is
estimated from those same $k$ maps, absorbing $1/k$ of the query variance. The bias is visible
directly --- the naive query-specific share is $11.3\%$, $22.0\%$ and $27.4\%$ for pages carrying
two, three and four queries --- so it also depends on how a sample is composed, which is why the
naive figures quoted in this section differ slightly. Correcting makes the estimate independent of
$k$, giving $35\%$ for all $k\ge3$.}

The query-specific third is real and lands where it should: on a calendar page, the query ``what
month and year are mentioned'' peaks on the text \textsc{July 2002}, while ``mention the page
number at top right corner'' peaks on \textsc{Page 11} instead. Because AGREE's loss is a cosine
against the \emph{whole} map, it is nonetheless computed mostly over the part that does not vary
with the query.

AGREE also contains a correction we do not use: dividing the query map by attention from a
query-agnostic prompt. In their pipeline that division is tied to the answer-token readout, their
weakest configuration, while the query-token readout we adopt --- their best --- uses the raw map.
We extracted query-agnostic maps for all 38{,}866 pages to see what dividing changes: on that
sample the naive query-specific share rises from $21\%$ to $29\%$ and the same-page cosine falls
from $0.68$ to $0.59$. Both are on the naive estimator, whose level moves with the sample, so read
the eight-point rise rather than the endpoints.

Our maps condition the teacher on the query alone, as AGREE's paper describes; AGREE's released extraction script also appends the gold answer, which we do not. We did not train a variant that changes any of this, so these are properties of the maps rather than of the result.

\paragraph{Relation to AGREE's reported results.} AGREE reports ViDoRe~v2 nDCG@5 rising from $58.59$ to $61.54$ for a ColQwen2.5 student and from $57.14$ to $57.75$ for a PaliGemma one \citep{agree}, with the backbone, training set, 7B query-token teacher, cosine loss and $\lambda{=}0.1$ we use, but under its own recipe: batch 128, learning rate $10^{-4}$, and a softplus loss on the hardest in-batch negative rather than InfoNCE at $T{=}0.02$. Those are single runs, and more than half of the ColQwen2.5 average gain comes from one of the four datasets (ESG human-labelled, $+7.0$). We kept our shared recipe so that all four signals are compared under identical conditions and did not run theirs, so the gap between their $+2.95$ and our $-0.1$ (per seed $-2.1$ to $+1.3$, Table~\ref{tab:seeds_e3}) may come from the recipe, from seed variation or from both, and we cannot separate them.

\section{Hyperparameter sweeps for attention and descriptions}
\label{app:sweeps}

\begin{table}[H]
\centering
\small
\begin{tabular}{ll rrr c l}
\toprule
arm & setting & \multicolumn{3}{c}{change vs.\ baseline} & seeds & verdict \\
\cmidrule(lr){3-5}
& & v1 & v2 & v3 & & \\
\midrule
A \; $\lambda_a$ & $0.05$ & $-0.2$ & $-0.1$ & $+0.7$ & 3 & \flips \\
 & $0.1$ \;(AGREE's paper) & $-0.1$ & $-0.6$ & $-0.4$ & 3 & \flips \\
 & $0.2$ & $+0.2$ & $-0.6$ & $+1.3$ & 3 & \flips \\
\addlinespace[3pt]
D \; $\tau_d$ & $0.01$ (ReAlign's code) & $-0.3$ & $-1.9$ & $-3.2$ & 3 & \consistneg \\
 & $0.05$ & $+0.4$ & $+2.5$ & $+1.0$ & 3 & \consistpos \\
 & $0.1$ & $+0.2$ & $-0.3$ & $-0.6$ & 3 & \flips \\
\addlinespace[3pt]
\bottomrule
\end{tabular}
\caption{Sweeps of the one hyperparameter that most controls each positive-level signal, at the one-epoch recipe (changes against a seed-matched baseline, nDCG@5). $\lambda_a$ weights the attention loss; $\tau_d$ is the temperature of the description KL. Each arm's marked row is the value its authors used --- $\lambda_a{=}0.1$ from AGREE's paper, $\tau_d{=}0.01$ from ReAlign's released training scripts, whose paper states no temperature. Table~\ref{tab:main} reports D at $\tau_d{=}0.05$.}
\label{tab:sweeps}
\end{table}
 Table~\ref{tab:sweeps} reports both sweeps in full.

Neither A nor D is reported solely at the value its authors used. $\lambda_a{=}0.1$ is AGREE's published setting; $\tau_d{=}0.01$ is the value in ReAlign's released training scripts, whose paper states no temperature. We swept each on the one hyperparameter
controlling how much its term can influence training, everything else fixed. For attention that is
the loss weight $\lambda_a$. For descriptions it is the KL temperature $\tau_d$ rather than
$\lambda_d$: at ReAlign's temperature the term's gradient comes from a few examples
(Appendix~\ref{app:repro}), and scaling it would mostly scale those.

The sweeps found opposite things. For descriptions the sweep changed the answer: ReAlign's released
temperature is negative at every seed, $\tau_d{=}0.05$ positive at every seed, and $\tau_d{=}0.1$
flips. Had we run only their value we would have concluded that descriptions do not help, when
what we had measured was a term concentrated on a few examples. For attention every setting flips sign
across its three seeds, so the null result does not depend on the weight; A is reported at its
published weight. Both sweeps were scored on the evaluation suites the main comparison uses, so
$\tau_d{=}0.05$ is a value selected on the test sets it is reported against, and D's figure should
be read as exploratory.

N was run at one configuration. S's temperature was changed once, from $T{=}1$ to $0.02$, after
the entropy diagnostic of Appendix~\ref{app:repro}; no other value was tried. The asymmetry still
runs against our conclusion: the arms we searched are the ones that helped least.

\section{Every seed of the main arms}
\label{app:perseed}

\begin{table}[H]
\centering
\small
\begin{tabular}{llccc}
\toprule
recipe & signal & seed 0 & seed 1 & seed 2 \\
& & \multicolumn{3}{c}{v1 / v2 / v3 nDCG@5} \\
\midrule
\emph{one epoch} & baseline & 86.3/49.0/42.8 & 87.0/51.7/45.4 & 87.6/56.7/49.1 \\
 & N & 89.1/62.2/51.7 & 89.2/63.2/51.7 & 89.5/61.3/52.0 \\
 & S & 88.9/63.3/53.1 & 88.7/62.5/51.9 & 88.7/63.4/52.7 \\
 & A & 86.1/49.0/43.3 & 86.7/53.0/45.1 & 87.7/53.7/47.9 \\
 & D & 86.9/52.3/45.1 & 87.3/55.6/45.8 & 87.8/57.0/49.4 \\
 & N$+$A & 89.4/62.2/51.7 & 89.4/61.1/51.4 & 89.6/61.4/51.2 \\
 & A$+$D & 86.8/53.7/42.7 & 87.0/54.1/44.3 & 87.8/51.1/45.4 \\
\addlinespace[3pt]
\emph{three epochs} & baseline & 87.9/55.3/47.3 & 88.3/55.0/47.5 & 88.5/55.3/47.5 \\
 & N & 89.8/63.4/52.0 & 89.3/61.9/51.7 & 89.6/62.5/52.2 \\
 & S & 89.0/61.4/52.2 & 89.0/63.6/52.6 & 88.8/64.0/53.3 \\
 & A & 88.8/55.7/47.6 & 88.0/53.0/46.0 & 88.5/56.6/47.8 \\
 & D & 88.9/57.4/49.4 & 88.5/57.7/47.6 & 88.4/58.2/48.6 \\
 & N$+$A & 88.9/61.6/51.3 & 89.0/62.9/51.1 & -- \\
\addlinespace[3pt]
\bottomrule
\end{tabular}
\caption{Every seed of the four signals, the baseline and their combinations, as absolute nDCG@5 on the three ViDoRe suites. The seed-mean of these numbers is what Table~\ref{tab:main} reports, and the spread between them is why we report per-seed ranges alongside seed means rather than an interval from a single run. S and D are at their corrected temperatures; \texttt{--} marks a combination we did not run.}
\label{tab:perseed}
\end{table}

Table~\ref{tab:perseed} gives every seed of the four signals, the baseline and their combinations,
so the seed-means in Table~\ref{tab:main} can be checked against the runs they average. At one
epoch the InfoNCE-only baseline alone scores 49.0, 51.7 and 56.7 on ViDoRe~v2 --- a 7.7-point
spread with nothing changed but the seed.

Tables~\ref{tab:seeds_e1} and~\ref{tab:seeds_e3} give the per-seed paired deltas at each
recipe, and Table~\ref{tab:seedaxis} decomposes the one-epoch spread by which source of
randomness was varied (\S\ref{sec:setup}).

\begin{table}[H]
\centering
\small
\begin{tabular}{llrrrrl}
\toprule
arm & suite & seed 0 & seed 1 & seed 2 & mean & verdict \\
\midrule
N & v1 & $+2.8$ & $+2.2$ & $+1.9$ & $+2.3$ & \consistpos \\
 & v2 & $+13.2$ & $+11.4$ & $+4.6$ & $+9.8$ & \consistpos \\
 & v3 & $+8.9$ & $+6.3$ & $+2.9$ & $+6.0$ & \consistpos \\
\addlinespace[2pt]
S & v1 & $+2.6$ & $+1.7$ & $+1.1$ & $+1.8$ & \consistpos \\
 & v2 & $+14.3$ & $+10.8$ & $+6.7$ & $+10.6$ & \consistpos \\
 & v3 & $+10.3$ & $+6.5$ & $+3.6$ & $+6.8$ & \consistpos \\
\addlinespace[2pt]
A & v1 & $-0.3$ & $-0.3$ & $+0.2$ & $-0.1$ & \flips \\
 & v2 & $0.0$ & $+1.3$ & $-3.0$ & $-0.6$ & \flips \\
 & v3 & $+0.5$ & $-0.3$ & $-1.2$ & $-0.4$ & \flips \\
\addlinespace[2pt]
D & v1 & $+0.6$ & $+0.3$ & $+0.2$ & $+0.4$ & \consistpos \\
 & v2 & $+3.4$ & $+3.9$ & $+0.4$ & $+2.5$ & \consistpos \\
 & v3 & $+2.3$ & $+0.4$ & $+0.3$ & $+1.0$ & \consistpos \\
\addlinespace[2pt]
N$+$A & v1 & $+3.0$ & $+2.4$ & $+2.0$ & $+2.5$ & \consistpos \\
 & v2 & $+13.2$ & $+9.4$ & $+4.8$ & $+9.1$ & \consistpos \\
 & v3 & $+8.9$ & $+6.0$ & $+2.1$ & $+5.7$ & \consistpos \\
\addlinespace[2pt]
A$+$D & v1 & $+0.5$ & $+0.1$ & $+0.2$ & $+0.2$ & \consistpos \\
 & v2 & $+4.7$ & $+2.3$ & $-5.6$ & $+0.5$ & \flips \\
 & v3 & $-0.1$ & $-1.1$ & $-3.8$ & $-1.7$ & \consistneg \\
\addlinespace[2pt]
\bottomrule
\end{tabular}
\caption{Per-seed paired deltas against the seed-matched InfoNCE-only baseline at the one-epoch recipe (nDCG@5$\times$100), macro-averaged over the datasets shared by both runs. S and D are the settings the paper reports ($T{=}0.02$ and $\tau_d{=}0.05$), as in Table~\ref{tab:main}; the published settings are swept separately in Table~\ref{tab:sweeps}. Table~\ref{tab:main} reports the three-epoch recipe, not this one.}
\label{tab:seeds_e1}
\end{table}

\begin{table}[H]
\centering
\small
\begin{tabular}{llrrrrl}
\toprule
arm & suite & seed 0 & seed 1 & seed 2 & mean & verdict \\
\midrule
N & v1 & $+1.9$ & $+1.0$ & $+1.1$ & $+1.4$ & \consistpos \\
 & v2 & $+8.1$ & $+6.9$ & $+7.2$ & $+7.4$ & \consistpos \\
 & v3 & $+4.6$ & $+4.2$ & $+4.7$ & $+4.5$ & \consistpos \\
\addlinespace[2pt]
S & v1 & $+1.1$ & $+0.7$ & $+0.4$ & $+0.7$ & \consistpos \\
 & v2 & $+6.1$ & $+8.6$ & $+8.7$ & $+7.8$ & \consistpos \\
 & v3 & $+4.9$ & $+5.1$ & $+5.9$ & $+5.3$ & \consistpos \\
\addlinespace[2pt]
A & v1 & $+0.9$ & $-0.3$ & $+0.0$ & $+0.2$ & \flips \\
 & v2 & $+0.4$ & $-2.1$ & $+1.3$ & $-0.1$ & \flips \\
 & v3 & $+0.2$ & $-1.5$ & $+0.3$ & $-0.3$ & \flips \\
\addlinespace[2pt]
D & v1 & $+1.1$ & $+0.3$ & $0.0$ & $+0.4$ & \flips \\
 & v2 & $+2.1$ & $+2.7$ & $+2.9$ & $+2.6$ & \consistpos \\
 & v3 & $+2.1$ & $+0.1$ & $+1.1$ & $+1.1$ & \consistpos \\
\addlinespace[2pt]
N$+$A & v1 & $+1.0$ & $+0.8$ & -- & $+0.9$ & \consistpos \\
 & v2 & $+6.3$ & $+7.8$ & -- & $+7.1$ & \consistpos \\
 & v3 & $+4.0$ & $+3.7$ & -- & $+3.8$ & \consistpos \\
\addlinespace[2pt]
\bottomrule
\end{tabular}
\caption{Per-seed paired deltas at the three-epoch recipe, our primary operating point (nDCG@5$\times$100). Conventions as in Table~\ref{tab:seeds_e1}. These are the runs averaged into Table~\ref{tab:main}.}
\label{tab:seeds_e3}
\end{table}

\begin{table}[H]
\centering
\small
\begin{tabular}{lrrr}
\toprule
run & v1 & v2 & v3 \\
\midrule
baseline & 86.3 & 49.0 & 45.1 \\
\quad batch order only & 85.7 & 52.6 & 46.7 \\
\quad projection init only & 86.3 & 52.2 & 47.0 \\
\quad LoRA init only & 86.9 & 46.5 & 47.3 \\
\quad control (identical rerun) & 86.3 & 49.0 & 45.1 \\
\midrule
A reference & 86.1 & 49.0 & 45.7 \\
\quad batch order only & 87.3 & 55.9 & 51.4 \\
\quad projection init only & 87.2 & 55.3 & 50.1 \\
\quad LoRA init only & 87.3 & 48.0 & 46.4 \\
\bottomrule
\end{tabular}
\caption{One-knob seed ablations at the one-epoch recipe. Changing only the batch order, or only the projection initialisation, moves ViDoRe~v2 as much as a nominally successful auxiliary loss. Absolute nDCG@5$\times$100. Three of the six runs are missing two ViDoRe~v3 sets, whose evaluation was truncated; every row is therefore averaged over the sets all six share (v1~10, v2~4, v3~6), so the columns are comparable to each other rather than to Table~\ref{tab:main}.}
\label{tab:seedaxis}
\end{table}

\begin{table}[H]
\centering
\small
\setlength{\tabcolsep}{4pt}
\begin{tabular}{lccc}
\toprule
arm & seed 0 & seed 1 & seed 2 \\
& \multicolumn{3}{c}{v1 / v2 / v3 nDCG@5} \\
\midrule
\multicolumn{4}{l}{\emph{baseline}} \\
InfoNCE only & 87.9/55.3/47.3 & 88.3/55.0/47.5 & 88.5/55.3/47.5 \\
\addlinespace[3pt]
\multicolumn{4}{l}{\emph{teacher-free rules (Table~\ref{tab:contributes})}} \\
N-mined, ranks 1--4 & 88.5/56.7/46.1 & 87.9/53.2/44.6 & -- \\
N-rank, ranks 4--20 & 88.5/54.5/46.8 & 88.6/56.7/46.1 & -- \\
N-band, ranks 21--100 & 89.0/59.0/50.0 & 89.2/56.3/48.7 & -- \\
N-perc, $\alpha{=}0.95$ & 89.8/57.8/49.3 & 89.9/59.5/49.6 & 90.3/58.3/49.8 \\
N-perc, $\alpha{=}0.864$ & 89.3/58.4/50.1 & 89.4/58.6/50.3 & -- \\
N, teacher-judged & 89.8/63.4/52.0 & 89.3/61.9/51.7 & 89.6/62.5/52.2 \\
\addlinespace[3pt]
\multicolumn{4}{l}{\emph{S's targets (Table~\ref{tab:spartition})}} \\
S, teacher ratings & 89.0/61.4/52.2 & 89.0/63.6/52.6 & 88.8/64.0/53.3 \\
flat target, no teacher & 88.0/53.5/44.4 & 86.8/50.7/38.4 & -- \\
S-masked, constant rows dropped & 87.4/66.0/53.6 & 87.6/66.4/53.8 & 88.2/64.7/54.4 \\
partition only, gradation dropped & 89.0/60.2/53.7 & 88.9/61.3/52.4 & -- \\
\addlinespace[3pt]
\bottomrule
\end{tabular}
\caption{Every seed of the arms that Tables~\ref{tab:contributes} and~\ref{tab:spartition} report as seed means, as absolute nDCG@5. Those tables give changes against a baseline trained with the same seed, which is the top row here, so every delta they quote can be recomputed column by column, to within the rounding shown. \texttt{--} marks a seed we did not run: the two arms here with three seeds are $\alpha{=}0.95$ and N, and S and S-masked are the two S arms with three. N and S repeat Table~\ref{tab:perseed} so that each comparison can be read off one table.}
\label{tab:perseedarms}
\end{table}

Table~\ref{tab:perseedarms} does the same for the arms that \S\ref{sec:contributes} and
\S\ref{sec:anatomy} report only as seed means: the five teacher-free selection rules and the
four targets S was trained against. One comparison in the main text crosses seed counts, and the
table makes it checkable. The $+1.7$ decomposition on v3 subtracts the two-seed mean of the
rate-matched threshold rule ($+2.8$) from N's three-seed mean ($+4.5$); on v2 both threshold settings give $+3.3$, so the same $+4.1$ holds at three matched seeds, N gaining $8.1$, $6.9$ and $7.2$ against the
$\alpha{=}0.95$ rule's $2.5$, $4.5$ and $3.0$. Read at three matched seeds on v3 instead,
against $\alpha{=}0.95$'s $+2.1$, the margin would be $+2.4$; we quote the smaller figure.

\section{Cost, in detail}
\label{app:cost}

\begin{table}[H]
\centering
\small
\begin{tabular}{lllrrr}
\toprule
signal & access & teacher work & \multicolumn{1}{r}{annot.} & train & total \\
\midrule
\emph{candidate mining} & --- & jina-embeddings-v4 over 38{,}866 pages & 9 & --- & \emph{shared by N, S} \\
\midrule
A (attention) & white-box & 1 map / positive & 24$^{\dagger}$ & $+$0 & \textbf{24} \\
D (descriptions) & black-box & 194.2M teacher tokens & $\approx$110 & $+$1.5 & \textbf{$\approx$112} \\
N (hard negatives) & black-box & 3.31M scored pairs & 37$\,+\,$9 & $\mathbf{+91}$ & \textbf{$\approx$137} \\
S (score distillation) & black-box & same pool, same pass & 37$\,+\,$9 & $+$91 & \textbf{$\approx$137} \\
\bottomrule
\end{tabular}
\caption{Measured cost of adopting each signal, in GPU-hours: annotating the full 118{,}195-pair training set, plus the \emph{extra} training cost of one three-epoch run relative to the InfoNCE-only baseline (23.9 GPU-h). Injecting $m{=}4$ documents per query makes a step $4.8\times$ more expensive, so N's marginal training cost (${+}91$) exceeds its entire annotation cost. Attention is free at training time. N and S require the same preparation --- the same candidate pools and the same single teacher pass --- so they cost the same, and the choice between them is not a cost decision. Candidate mining is listed separately because it precedes the teacher entirely: N and S both need it, A and D do not. $^{\dagger}$Attention is the one arm run with the 7B teacher rather than the 72B, because that is the configuration its authors select; at 72B the same extraction would cost 130--280 GPU-hours.}
\label{tab:cost}
\end{table}

Table~\ref{tab:cost} gives the totals. All figures are H100 GPU-hours, wall-clock $\times$
devices, for processing the full 118{,}195-pair training set once.

\textbf{Mining} (N, S): encoding 38{,}866 pages 5.9, encoding queries 0.5, single-vector retrieval
to depth 5{,}000 plus MaxSim reranking 2.2 --- about 9 in total. A and D never use it.
\textbf{Scoring} (N, S): 37 with the 72B teacher across two shards, 3.31M scored pairs. Ordering
the queue by page rather than query gave a 5.5$\times$ speedup, since a page recurs in many pools
and its image encode is reused. \textbf{Attention} (A): 24 with the 7B teacher, one map per
positive; the only signal needing white-box access, which is a harder constraint than the hours.
This is the one figure not comparable across arms, since every other signal uses the 72B teacher;
we began on the 72B and switched to AGREE's best-reported configuration, 7B with the query-token readout; AGREE reports no 72B query-token run, and at the answer-token readout its 72B beats its 7B. Two effects
separate them. The 72B is an order of magnitude larger, and attention extraction cannot use the
vLLM server the scoring and description passes run on, because reading attention weights requires
the unfused eager path, which materialises a $[\text{heads}, \text{seq}, \text{seq}]$ tensor per
layer. Measured throughput was 1.58 pages/s for the 7B against 0.12--0.25 for the 72B --- a
6--13$\times$ spread, the width of it reflecting different batch and resolution settings across
the partial 72B runs. The full training set at the 72B rate is therefore 130--280 GPU-hours. That
range, not a point estimate, is what the evidence supports. \textbf{Descriptions} (D):
${\approx}110$, the most expensive by far, because generation is sequential where scoring emits a
few tokens --- 194.2M teacher tokens, 175.8M prompt and 18.3M generated.

\textbf{Training.} On identical hardware a step costs 7.8\,s for the baseline, 7.8\,s with
attention, 8.3\,s with descriptions and 37.4\,s for any arm appending four documents --- those
encode five pages per query where the others encode one. Over three epochs that is ${+}91$
GPU-hours against the baseline's 23.9, so for N the training bill is roughly twice the annotation
bill, and it recurs on every retrain.

\section{The human validation study}
\label{app:study}

\paragraph{Sampling.} Fifty
queries were drawn at random from the \emph{query-complete} pool ---
those where the labelled positive and all four top-ranked mined candidates carry a teacher score
--- stratified so that the natural rate of queries carrying a candidate at exactly $0.5$ was
preserved rather than enriched. Each query contributes four pages: twenty queries show the
labelled positive plus its top three mined candidates, the other thirty show the top four
candidates and no positive. Uniform page counts mean block size leaks nothing, and the positive is
present on a minority of queries so its presence is hard to infer. The $2.04$ of \S\ref{sec:labels} counts each annotator's view of a query as one observation rather than merging the two annotators' grades per page, which would need a tie-break rule the study deliberately does not have. A view counts only where none of its four grades is \emph{cannot judge}: nine of the sixty views over these queries carry at least one and are dropped, which removes one query outright and leaves $51$ views over $29$ queries.

\paragraph{Assignment.} The fifty queries were cut into ten blocks of five, stratified so each
block carries its share of the hedge-bearing queries and exactly two positive-bearing ones.
Annotator $i$ graded blocks $i$ and $i{+}1 \bmod 10$. Every page is therefore judged twice by
different people, every annotator overlaps with two others, and the ten pairings that result are
ten distinct pairs of people rather than five fixed couples --- so $\kappa$ averages over raters
instead of being a property of one partnership. Each annotator saw $40$ pages, about $25$ minutes
of work.

\paragraph{Rubric.} The UMBRELA four-point scale \citep{umbrela}: \textbf{3} fully answers the
query, \textbf{2} gives part of the answer, \textbf{1} on topic but does not answer, \textbf{0}
unrelated, plus \textbf{x} for cannot judge. The instructions named the $1$ versus $2$ boundary as
the one that matters and gave the test for it --- \emph{could you write down some part of the true
answer, citing only this page?} --- because that boundary is where the binary cut at ${\geq}2$
falls. Annotators were also told that a keyword match is not relevance, since each page came with
a machine transcript for searching.

\paragraph{Blinding.} Teacher scores lived in a file no viewer loaded, the positive was never
marked, and page order was shuffled independently per annotator so a query's pages were never
adjacent. The \emph{specific}/\emph{broad} classification of all fifty queries was written by hand by an author who could not see the teacher's scores, and frozen before any grade was read; the analysis refuses to report that split if any query is
unlabelled.

\begin{table}[H]
\centering
\small
\begin{minipage}[t]{0.56\textwidth}
\centering
\begin{tabular}{lrrrrr}
\toprule
rater & med.\ s & \emph{x} & rel.\ & vs.\ teacher & gold \\
\midrule
v01 & 41 & 0.0\% & 47\% & 75\% & 3/4 \\
v02 & 36 & 2.5\% & 11\% & 80\% & 4/4 \\
v03 & 30 & 10.0\% & 38\% & 69\% & 2/4 \\
v04 & 34 & 0.0\% & 67\% & 81\% & 4/4 \\
v05 & 8 & 12.5\% & 81\% & 72\% & 2/3 \\
v06 & 25 & 0.0\% & 50\% & 81\% & 2/4 \\
v07 & 18 & 2.5\% & 51\% & 83\% & 4/4 \\
v08 & 27 & 17.5\% & 45\% & 72\% & 4/4 \\
v09 & 40 & 12.5\% & 35\% & 77\% & 3/4 \\
v10 & 19 & 0.0\% & 44\% & 78\% & 4/4 \\
\bottomrule
\end{tabular}
\end{minipage}\hfill
\begin{minipage}[t]{0.38\textwidth}
\centering
\begin{tabular}{lrr}
\toprule
pairing & $n$ & $\kappa$ \\
\midrule
v01--v02 & 17 & $0.85$ \\
v01--v10 & 18 & $0.87$ \\
v02--v03 & 14 & $0.00$ \\
v03--v04 & 18 & $0.33$ \\
v04--v05 & 15 & $-0.13$ \\
v05--v06 & 17 & $0.16$ \\
v06--v07 & 17 & $0.53$ \\
v07--v08 & 17 & $0.65$ \\
v08--v09 & 11 & $0.44$ \\
v09--v10 & 17 & $0.39$ \\
\bottomrule
\end{tabular}
\end{minipage}
\caption{Per annotator (left) and per ring pairing (right). \emph{med.\ s} is the median seconds per judgement, \emph{x} the cannot-judge rate, \emph{rel.} the share of mined candidates the annotator called relevant, \emph{vs.\ teacher} their raw agreement with the teacher's binary decision, and \emph{gold} how many of the labelled positives they saw they judged relevant. The relevant-rate column spans 11\% to 81\%; since each annotator graded a different block of queries, that spread mixes strictness with the queries drawn (Appendix~\ref{app:study}). Human--human $\kappa$ is 0.41, and agreement with the teacher is nearly flat across all ten. No annotator is excluded; the prose reports a leave-one-out check instead.}
\label{tab:studyraters}
\end{table}
 Table~\ref{tab:studyraters} gives the per-annotator breakdown.

\paragraph{No annotator is excluded.} One annotator graded much faster than the rest (median $8$
seconds against $29$) and returned the highest relevant-rate at $81\%$; another returned the lowest
at $11\%$. Each graded a different block of ten queries, so a marginal rate mixes how strictly an
annotator judged with which queries they drew, and neither figure measures strictness on its own.
Both are more than three median-absolute-deviations from the median relevant-rate, in opposite
directions, and only the first would have been noticed by inspecting agreement --- the strict one agrees well with the teacher and with one of its two ring partners ($\kappa$ $0.85$; $0.00$ with the other). Selecting an exclusion rule after seeing which
annotator looks bad is the failure the frozen analysis exists to prevent, so we exclude nobody and
report robustness directly: dropping each annotator in turn moves precision of protection within
$68$--$73\%$, the miss rate within $11$--$17\%$, relevant-pages-per-query within $1.93$--$2.24$,
and leaves the specific/broad intervals disjoint in all ten cases. The largest shift any single
annotator causes in any headline figure is four points.

\section{Contamination by stratum}
\label{app:strata}

\S\ref{sec:labels} reports the severe stratum's share and back-fill rate as text.
Table~\ref{tab:strata} gives the full split.

\begin{table}[H]
\centering
\small
\begin{tabular}{llrrr}
\toprule
stratum & $c(q)$ & share of queries & back-fill fires & N's negatives rated relevant \\
\midrule
clean & $0$ & 18.9\% & 0.0\% & 0.00 of 4 \\
moderate & 1--2 & 25.4\% & 0.3\% & 0.00 of 4 \\
\bfseries severe & \bfseries 3--4 & \bfseries 55.7\% & \bfseries 45.1\% & \bfseries 1.44 of 4 \\
\bottomrule
\end{tabular}
\caption{Where the filter is incomplete. $c(q)$ counts how many of the four top-ranked non-gold candidates the teacher rates relevant. On the severe stratum the filter runs out of below-threshold candidates for 45\% of queries and back-fills with pages it rated relevant itself, so it is least complete exactly where there is most to remove. Pooled over all strata, 20.0\% of the negatives N trains on are teacher-rated relevant, reproducing the corpus-wide figure in Table~\ref{tab:contributes} from the selection rule directly.}
\label{tab:strata}
\end{table}

Two consequences travel with this table. Filtering is close to a no-op on the clean stratum ---
with no top-four candidate above threshold there is nothing to remove --- so an arm restricted to
clean queries would be vacuous by construction. And because the severe stratum is the one that
back-fills, any attempt to localise N's gain by stratum is confounded: the stratum with the most
contamination to remove is also the stratum where removal is least complete. That is the
experiment the next appendix reports.

\section{Where the filter is applied}
\label{app:localisation}

\S\ref{sec:contributes} shows that the teacher's per-query filter beats every teacher-free rule we tried on both primary suites, and Table~\ref{tab:strata} shows the contamination it removes is concentrated in one
stratum. A natural follow-up is whether the gain can be localised: if most of the contamination
sits on the severe queries, does treating only those recover most of N?

We ran that at a fixed budget. Four arms each apply the filter to one manifest of $15{,}813$
queries --- the moderate stratum, the severe stratum, the severe queries the filter satisfies
without back-filling, and a random draw from the eligible population, source-matched to its
stratum mix --- and leave every other query on the unfiltered top-4. Every arm therefore sees the
whole corpus and appends four candidates per query; only which queries the teacher is consulted
about differs. Count-matching cannot equalise the \emph{dose}, the number of teacher-rated
relevant candidates actually removed per treated query, so we report it beside each arm.

\begin{table}[H]
\centering
\small
\begin{tabular}{llllll}
\toprule
& & \multicolumn{2}{c}{ViDoRe~2} & \multicolumn{2}{c}{ViDoRe~3} \\
\cmidrule(lr){3-4}\cmidrule(lr){5-6}
treated queries & dose & per seed & verdict & per seed & verdict \\
\midrule
$c(q){=}1,2$ (moderate) & 1.47 & $-0.45$ / $+0.19$ & \flips & $-0.60$ / $-1.17$ & \consistneg \\
random eligible & 2.06 & $+2.58$ / $+0.22$ & \consistpos & $+0.26$ / $+0.50$ & \consistpos \\
$c(q){=}3,4$ (severe) & 2.44 & $+1.24$ / $+1.71$ & \consistpos & $+0.35$ / $-0.11$ & \flips \\
\quad severe, no back-fill & -- & $+0.94$ / $+1.47$ & \consistpos & $-0.86$ / $-0.11$ & \consistneg \\
\bottomrule
\end{tabular}
\caption{Where the filter is applied, at a fixed budget of $15{,}813$ treated queries; every other query keeps the unfiltered top-4. \emph{dose} is the mean number of teacher-rated relevant candidates the filter removes per treated query. Per-seed changes against a seed-matched baseline (nDCG@5), two seeds.}
\label{tab:localisation}
\end{table}
 Table~\ref{tab:localisation} reports the four arms.

The experiment is inconclusive. Two findings point the way the contamination account predicts: treating the severe
stratum helps, and treating the moderate one does not, in line with their doses. Two do not: a
random eligible draw at a lower dose does as well as the severe stratum on ViDoRe~v2, and on v3
the random draw is the only arm consistently positive on v3.

The precision is the problem. Every mean effect here is under $1.5$ points at two seeds, and the random
draw's two v2 seeds alone span $2.4$ points, so it cannot be placed against the arms it is meant
to control for. We therefore draw no conclusion about whether targeting contamination matters,
beyond the negative one recorded with the strata table (Appendix~\ref{app:strata}): an arm
restricted to the clean stratum would be vacuous, because there is nothing there to remove.

One result in the table is legible on its own. Removing back-fill from the severe arm --- so the
filter only treats queries where it can find four candidates below its threshold without falling
back on pages it rated relevant --- costs little. N's own back-fill fires on $45\%$ of severe
queries (Table~\ref{tab:strata}), and this is the closest we come to measuring what that costs:
not much.

\section{Per-dataset breakdown}
\label{app:tables}

Suite averages can hide dataset-level variability, so Table~\ref{tab:heatmap} gives the change on
every constituent of the two out-of-domain suites. N, S and D are positive on all twelve; the
gains vary about fourfold across the twelve datasets, but no sign moves.

\begin{table}[H]
\centering
\small
\begin{tabular}{lrrrr}
\toprule
dataset & N & S & D & A \\
\midrule
\multicolumn{5}{l}{\emph{ViDoRe~2}} \\
\quad biomedical lectures & \cellcolor{dvpos2}$+6.0$ & \cellcolor{dvpos2}$+6.7$ & \cellcolor{dvzero}$+1.5$ & \cellcolor{dvzero}$-0.1$ \\
\quad economics reports & \cellcolor{dvpos3}$+11.2$ & \cellcolor{dvpos3}$+11.1$ & \cellcolor{dvpos2}$+4.1$ & \cellcolor{dvpos1}$+2.3$ \\
\quad esg reports human labeled & \cellcolor{dvpos2}$+5.9$ & \cellcolor{dvpos3}$+8.9$ & \cellcolor{dvpos1}$+3.1$ & \cellcolor{dvneg1}$-1.9$ \\
\quad esg reports & \cellcolor{dvpos2}$+6.5$ & \cellcolor{dvpos2}$+4.5$ & \cellcolor{dvpos1}$+1.6$ & \cellcolor{dvzero}$-0.9$ \\
\multicolumn{5}{l}{\emph{ViDoRe~3}} \\
\quad computer science & \cellcolor{dvpos2}$+4.9$ & \cellcolor{dvpos2}$+5.7$ & \cellcolor{dvpos1}$+1.8$ & \cellcolor{dvzero}$-0.4$ \\
\quad energy & \cellcolor{dvpos2}$+4.6$ & \cellcolor{dvpos2}$+4.4$ & \cellcolor{dvzero}$+0.9$ & \cellcolor{dvzero}$+0.5$ \\
\quad finance en & \cellcolor{dvpos1}$+3.2$ & \cellcolor{dvpos2}$+5.3$ & \cellcolor{dvzero}$+0.6$ & \cellcolor{dvzero}$-0.3$ \\
\quad finance fr & \cellcolor{dvpos1}$+2.5$ & \cellcolor{dvpos1}$+3.0$ & \cellcolor{dvzero}$+0.3$ & \cellcolor{dvzero}$-0.8$ \\
\quad hr & \cellcolor{dvpos2}$+7.3$ & \cellcolor{dvpos2}$+7.3$ & \cellcolor{dvpos1}$+2.0$ & \cellcolor{dvzero}$+0.2$ \\
\quad industrial & \cellcolor{dvpos2}$+6.6$ & \cellcolor{dvpos2}$+6.3$ & \cellcolor{dvpos1}$+2.3$ & \cellcolor{dvneg1}$-1.7$ \\
\quad pharmaceuticals & \cellcolor{dvpos2}$+4.5$ & \cellcolor{dvpos2}$+7.1$ & \cellcolor{dvzero}$+0.7$ & \cellcolor{dvzero}$-0.7$ \\
\quad physics & \cellcolor{dvpos1}$+2.7$ & \cellcolor{dvpos1}$+3.1$ & \cellcolor{dvzero}$+0.4$ & \cellcolor{dvzero}$+0.5$ \\
\bottomrule
\end{tabular}
\caption{Change against a seed-matched baseline on every constituent of the two out-of-domain suites (nDCG@5, seed-means). N and S are positive on all 12 of them, D on all of them but by smaller margins, and A changes sign across datasets and never by more than about two points. So the suite averages are not carried by one subset --- the gains vary about fourfold across the twelve datasets for N and S (largest on economics reports, smallest on physics and French finance), but the sign never moves. Shading is a diverging scale with symmetric thresholds at $1.5$, $4$ and $8$ points; values are printed as well as shaded so the table reads without colour.}
\label{tab:heatmap}
\end{table}

\section{What the teacher's filter removes}
\label{app:falseneg}

Table~\ref{tab:negatives} gives the teacher's rating of mined candidates by rank band, which is
what motivates the teacher-free controls of \S\ref{sec:contributes}.

\begin{table}[H]
\centering
\small
\begin{tabular}{lrr}
\toprule
retriever rank (non-gold) & mean score & \% rated relevant \\
\midrule
1 & 0.645 & 70.1\% \\
2 & 0.570 & 62.4\% \\
3--4 & 0.519 & 57.0\% \\
5--9 & 0.466 & 51.1\% \\
10--20 & 0.410 & 45.1\% \\
21--100 & 0.326 & 35.7\% \\
\midrule
labelled gold page & 0.969 & 98.5\% \\
\bottomrule
\end{tabular}
\caption{What the teacher thinks of the retriever's candidates, by position among the non-gold candidates, rank~1 being the highest-ranked one --- the convention of \S\ref{sec:contributes}, and the one the negatives are selected in. Each row is the mean over \emph{candidates at that rank band}, so the rank-1 row is the first page any hard-negative miner would take: rated relevant for 70\% of queries. The filter routinely declines it, moving the selected negatives from median rank 2 to median rank 5. Whether those candidates are genuinely relevant is answered directly in \S\ref{sec:validation}, where annotators judge $2.04$ of a query's four top-ranked mined candidates relevant; \S\ref{sec:labels} gives the corpus-level account and Appendix~\ref{app:falseneg} shows what they look like.}
\label{tab:negatives}
\end{table}

\paragraph{How deep the filter reaches.} The teacher's per-query decision amounts to pushing
negatives deeper by an amount it chooses per query rather than by a fixed window.
\textbf{$22.9\%$} of the negatives N accepts come from beyond rank~10, while on the $18.9\%$ of
queries with no rated-relevant page in the top four it keeps all four of theirs there
(Table~\ref{tab:strata}) --- the depth it chooses varies with the query rather than being set
in advance. A fixed window applies one depth
to every query and cannot reproduce this: the window covering the teacher's own typical range
gains $0.4$ points, while the best window is a deeper one (\S\ref{sec:contributes}).
The teacher scores 28 candidates per query: the top 20, plus 8 drawn one per ten ranks from 21 to
100. Rates for ranks 1--20 are therefore exhaustive, while the rate for the 21--100 band is a stratified estimate over the eight sampled ranks.

\S\ref{sec:labels} argues that the top of a mined pool is contaminated because the corpus's
queries are \emph{corpus-generic}: many pages answer them. This appendix gives the source-level
evidence. The rejected candidates are unlabelled positives in the sense that matters for training
--- the page answers the query as posed, so pushing it away is a corrupted label --- and many are not artefacts of a vague question, since ``what is the date?'' has a perfectly checkable answer on
any given page (\S\ref{sec:labels}). What varies by source is how many pages carry such an answer:

\begin{center}\small
\begin{tabular}{lrr}
\toprule
source & queries & rated relevant \\
\midrule
DocVQA & 39{,}301 & 81\% \\
TAT-DQA & 13{,}188 & 83\% \\
PDF (synthetic) & 45{,}718 & 65\% \\
arXiv-QA & 9{,}949 & 62\% \\
Infographic-VQA & 10{,}038 & 43\% \\
\midrule
all & 118{,}194 & 70\% \\
\bottomrule
\end{tabular}
\end{center}

Contamination is therefore real, and highest in the form-reading sources, consistent with corpus-generic queries; and mining
hard negatives for such a query is ill-posed however good the retriever is. This is the boundary
condition \S\ref{sec:conclusion} states as a prediction: the value of a teacher filter depends on
how many pages a corpus's queries each admit.

\paragraph{What this looks like.} Four queries, with the teacher's score for each of the four
highest-ranked non-gold candidates. The first three are the failure mode; the fourth is what a
well-posed query looks like.

\begin{center}\small
\begin{tabular}{p{0.52\linewidth} cccc}
\toprule
query & r1 & r2 & r3 & r4 \\
\midrule
``What is the date?''                            & 1.00 & 1.00 & 1.00 & 1.00 \\
``To whom should the form be returned?''         & 1.00 & 1.00 & 1.00 & 1.00 \\
``What is Bert Shulimson's title?''              & 1.00 & 1.00 & 1.00 & 1.00 \\
\addlinespace[2pt]
``What are some applications of the CR36L process?'' & 0.00 & 0.00 & 0.00 & 0.00 \\
\bottomrule
\end{tabular}
\end{center}

For the first three, every candidate a miner would take is a page the teacher rates fully
relevant, so training on them as negatives teaches the model that relevant pages are wrong; for
the fourth they are all irrelevant and mining costs nothing. \textbf{Of the $115{,}698$ queries
whose four top-ranked non-gold candidates are all scored, $40\%$ have at least three of them rated
${\geq}0.9$.} The first three are DocVQA-style form-reading questions and the fourth asks about
specific content, which is the source effect above.

\section{Extended limitations}
\label{app:limitations}

\S\ref{sec:conclusion} states these in brief. This appendix gives each in full.

We study one student family (ColQwen2.5-3B), one training corpus, and one late-interaction
architecture. A larger student, a different backbone, or a single-vector retriever could reorder
the signals. That caveat is sharpest for descriptions, whose published form targets single-vector
students and which we had to port to MaxSim scoring. The attention arm uses a 7B teacher while the
score- and description-based arms use a 72B, because that is the configuration the attention
method's authors select. Matched teacher scale therefore holds across three of the four signals
but not the fourth.

We did not run a hyperparameter search. N was run at one configuration, and S's temperature was changed once after a target-entropy diagnostic (Appendix~\ref{app:repro}). Attention and description alignment were each swept over three values of the one
hyperparameter that most controls their influence on training, and descriptions are reported at
the swept value rather than the one in ReAlign's released code (\S\ref{sec:results}, Appendix~\ref{app:sweeps}).
That is a screen, not an optimisation: three points on one axis per arm, three seeds each. A signal that does not help here may still
be untuned on an axis we did not vary, and our own results show how much a single hyperparameter
can decide: at ReAlign's temperature, description alignment is below its baseline at every
seed (Table~\ref{tab:sweeps}). Our
coverage of the seed axis is similarly thin: two to three runs at three epochs and three at one
epoch. That is enough for a consistent sign and a consistent ordering of the mean gains, which is
what we claim, and for N and S those gains are several times the spread between seeds; it is not
enough for a precise effect size.

The teacher's judgements are audited against human ones in \S\ref{sec:validation}, but only
partly. Two gaps matter most here: the audit covers only the four top-ranked candidates, so every
negative N draws from below them was never judged --- at least one negative on the $81\%$ of
queries where a top-four candidate clears the threshold (Table~\ref{tab:strata}), and all four on
the $34\%$ where every one does; and the audit establishes that the teacher
agrees with human judgements about as well as they agree with each other, not that a filter with
better judgements would train a better retriever, which no measurement here addresses.

We also never train on a continuous teacher target. \S\ref{sec:anatomy} shows the rating readout
is close to binary while a yes/no logprob readout of the same teacher is not
(Appendix~\ref{app:teacherprompt}). So the coarseness we report is a property of how we elicited
the judgement, and we cannot say whether a genuinely graded target would have done better.
Re-annotating is cheap, but retraining both signals that consume the scores is not, and a
logprob-trained arm would in any case be a different experiment rather than the same one with a
better input; Appendix~\ref{app:teacherprompt} gives the cost.

N and S cost about $4.8\times$ as much per step as the baseline. The controls in
\S\ref{sec:contributes} hold that fixed for N --- they append four documents too, differing only
in which --- so N's margin over them is not an artefact of compute. S has a compute-matched teacher-free control of its own: the flat-target arm of
\S\ref{sec:anatomy} keeps its loss, its weight, its candidates and its forward passes and uses no
teacher, and loses $3.1$ points on v2. Its candidates enter only its KL term and never the InfoNCE
denominator, so its gain is not compute or exposure either. What we lack is an \emph{informative}
teacher-free target for S. Because S discards nothing, the false-negative mechanism that may explain N's advantage over a fixed rank rule need not apply to it, and a cheaper target --- the
retriever's own scores over the same candidates --- might recover part of its gain.
That control needs no new annotation, since the pools and those scores already exist, and we leave
it to future work.

We compare signals one at a time and try only two combinations. Training hard negatives and
attention grounding together leaves the result where hard negatives alone put it ($+7.1$ against
$+7.5$ on ViDoRe~v2 at matched seeds), so whether a positive-level and a candidate-level signal
can complement each other is left open; the other combination, attention with descriptions, was run at one epoch only and flips sign on v2 across its seeds (Table~\ref{tab:seeds_e1}). Nor do we combine the two \emph{candidate-level} signals
with each other: applied together they would each contribute their own four pages per query, so
the arm would no longer be matched to the others on candidate exposure or compute. Whether they
are complementary or redundant --- both draw on the same teacher judgements, one to select
negatives and one to shape a target --- is therefore also left open; in text retrieval, \citet{arhn} find filtering and relabelling LLM-judged negatives complementary. Finally, we measure
page-level retrieval only. A localisation signal could improve grounding --- in the sense the
attention method's own region-level evaluation captures --- without improving nDCG@5. Our results
for attention and descriptions concern retrieval quality, not grounding.

\end{document}